\documentclass[]{interact}
\usepackage{epstopdf}
\usepackage[caption=false]{subfig}

\usepackage[numbers,sort&compress]{natbib}

\usepackage{kpfonts}
\usepackage[table]{xcolor}
\usepackage{kpfonts}
\usepackage{amsmath,bm,hyperref,url,fullpage,mathtools,booktabs,parskip}
\usepackage{paralist,enumitem,todonotes,microtype}
\usepackage{graphicx}
\usepackage{url} 
\usepackage{algorithm}
\usepackage{algorithmicx}
\usepackage{algpseudocode}
\usepackage{verbatim}
\usepackage{multirow}
\usepackage{multicol}
\usepackage{arydshln}
\usepackage{color}
\usepackage{adjustbox}
\usepackage{amssymb}
\usepackage{amsmath}
\usepackage{mathtools}
\usepackage{bbm}
\usepackage{amsmath,bm,url,mathtools,booktabs,parskip}
\usepackage{graphicx}
\usepackage{placeins}
\usepackage{setspace}
\usepackage{appendix}

\usepackage{amsmath}
\usepackage{graphicx,psfrag,epsf}
\usepackage{enumerate}
\usepackage{natbib}

\usepackage{algpseudocode}

\providecommand{\tightlist}{%
  \setlength{\itemsep}{0pt}\setlength{\parskip}{0pt}}
\let\code=\texttt
\let\proglang=\textsf

\usepackage{float}
\hypersetup{
hidelinks,
colorlinks=false,
linkcolor=black,
citecolor=black
}

\bibpunct[, ]{[}{]}{,}{n}{,}{,}
\renewcommand\bibfont{\fontsize{10}{12}\selectfont}

\theoremstyle{plain}

\theoremstyle{definition}

\theoremstyle{remark}

\def\Phi{\phi}
\def\Theta{\theta}
\def\Psi{\psi}

\def\tilde#1{#1}

\begin{document}

\articletype{}

\title{DeepVARwT: Deep Learning for a VAR Model with Trend}

\author{
\name{Xixi Li\textsuperscript{a}\thanks{CONTACT Jingsong Yuan. Email: j.yuan@manchester.ac.uk} and Jingsong Yuan\textsuperscript{a}}
\affil{\textsuperscript{a}Department of Mathematics, University of Manchester, UK}
}

\maketitle

\begin{abstract}
Time series modelling and prediction is useful in many fields of application such as economics, finance and engineering.  
The vector autoregressive (VAR) model has been used to describe the dependence within and across multiple time series. This is a model for stationary time series,  which can be extended to allow the presence of a deterministic trend in each series. 
In this paper, we demonstrate a new approach that employs deep learning methodology for maximum likelihood estimation of the trend and the dependence structure at the same time. A Long Short-Term Memory (LSTM) network is used for this purpose. 
We provide a simulation study and applications to real data. In the simulation study, we use realistic trend functions estimated from real data and compare the estimates with true function/parameter values. In the real data applications, we compare the prediction performance of this model with state-of-the-art models in the literature.
\end{abstract}


\section{Introduction}\label{sec1}

\begin{figure}[h!]
	\centering
	\includegraphics[width=0.6\linewidth]{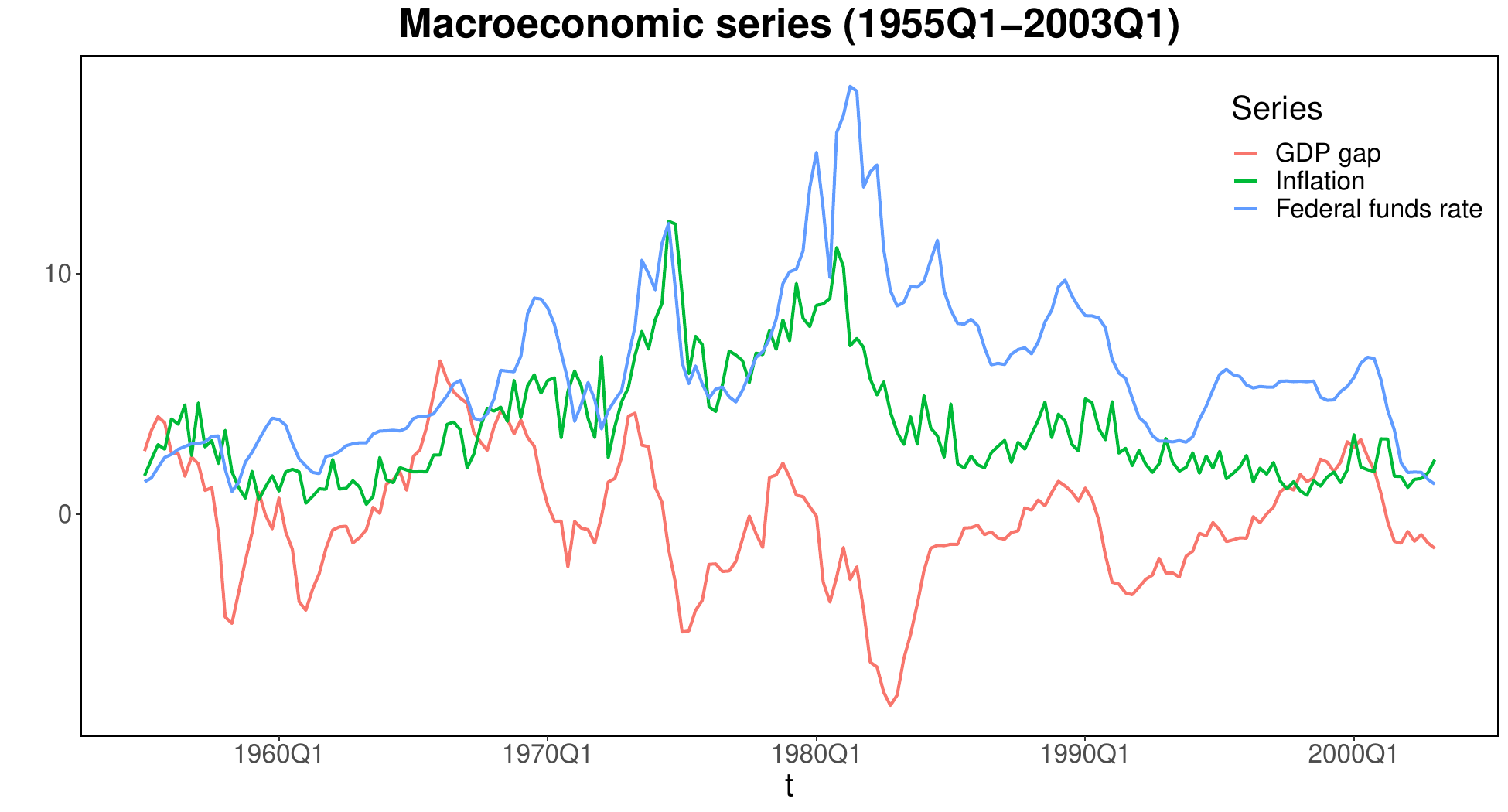}
			\vspace*{-5mm}
	\caption{US macroeconomic series spanning 1955Q1 to 2003Q1.} 
	\label{fig:real-data1}
\end{figure}

In practice, many time series exhibit nonstationary characteristics in the mean. For example, Fig.~\ref{fig:real-data1} shows 
three quarterly US macroeconomic series, namely GDP gap, inflation, and federal funds rate, as analysed by \cite{jorda2005estimation}. Each series is nonstationary as the mean is apparently not constant.  
More examples of time series with trends will be given in Sections~\ref{temp-sec} and~\ref{usmacro-bvarsv-sec}. 

A simple approach to detrending a time series is to difference it until it appears to be stationary. This is effective when the trend is a low order polynomial. However, the trend itself may be of interest, 
which can be estimated by smoothing the data {\color{black}as if the errors are i.i.d.}, using methods such as Kernel Smoothing \citep{wand1994kernel}, Locally Weighted Scatterplot Smoothing (Lowess), or Smoothing Splines, to name just a few. The series after removing the trend in each component can then be analysed by fitting a stationary model. Inference on model parameters will have to ignore errors in estimating the trend in this semi-parametric approach in two stages. {\color{black}It will be better to estimate the trend and dependence parameters at the same time and properly by the maximum likelihood principle{\color{black}, which we do in this paper.}}  

The vector autoregressive VAR($p$) model 
 \begin{equation}
 \label{var}
  \begin{aligned}
\mathbf{y}_{t}=A_{1} \mathbf{y}_{t-1}+A_{2} \mathbf{y}_{t-2}+\cdots+A_{p} \mathbf{y}_{t-p}+\bm{\varepsilon}_{t},\qquad   t=0, \pm 1, \pm 2,     \ldots, 
\end{aligned}
\end{equation}
is for stationary time series $\{\mathbf{y}_t\}$, where $A_1$,\ldots,$A_p$ are constant coefficient matrices, and 
$\{\bm{\varepsilon}_{t}\}$ is multivariate white noise. 
It can be extended to accommodate a polynomial trend in each series. If we assume the mean $\bm{\mu}_t$ of $\mathbf{y}_{t}$ consists of $\color{black}r$-th order polynomials, and $\{{\mathbf y}_t-\bm{\mu}_t\}$ satisfies the VAR model (\ref{var}), then a VAR with trend (VARwT) model can be written as 
\begin{equation}
\label{varx}
  \begin{aligned}  
\mathbf{y}_{t}=A_{1} \mathbf{y}_{t-1}+A_{2} \mathbf{y}_{t-2}+\cdots+A_{p} \mathbf{y}_{t-p}+C {\mathbf {\color{black}z}}_t+\bm{\varepsilon}_{t},\qquad   t=0, \pm 1, \pm 2,     \ldots, 
\end{aligned}
\end{equation}
where ${\mathbf {\color{black}z}}_t=(1, t, t^{2}, \ldots, t^{\color{black}r})'$ and $C$ is a matrix of constants. Both the trend and the dependence parameters can be estimated simultaneously using ordinary least squares \citep{vars}{\color{black}, which is equivalent to maximum {\em conditional} likelihood estimation assuming Gaussian errors \cite{lutkepohl2005new}.}

\begin{figure}[h!]
	\centering
 \includegraphics[width=0.55\linewidth]
 {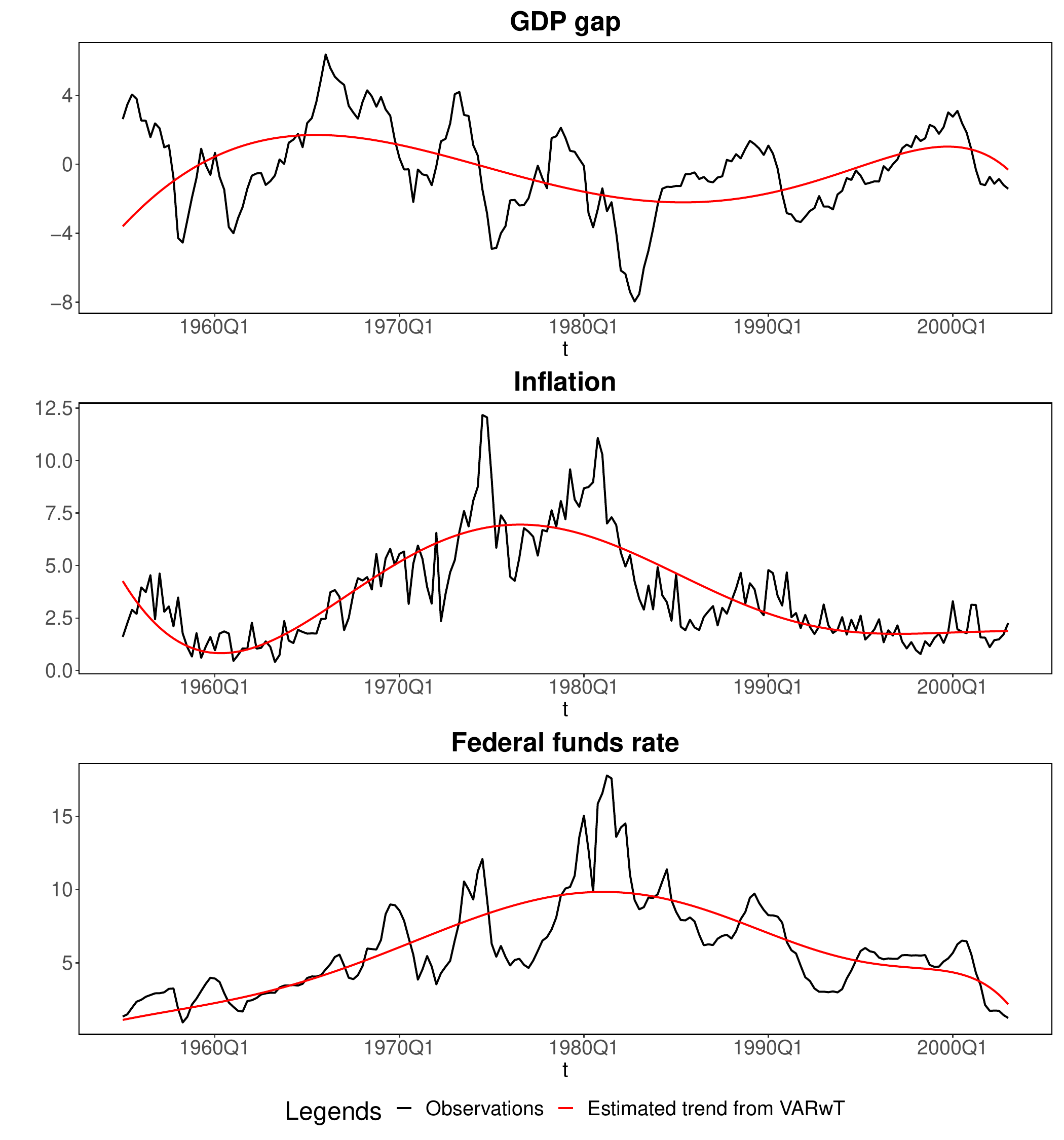}
			\vspace*{-8mm}
	\caption{US macroeconomic series and estimated polynomial trends (red lines).} 
	\label{fig:estimated-trend-us-macro}
\end{figure}

Fig.~\ref{fig:estimated-trend-us-macro} shows polynomial trends estimated together with VAR($4$) coefficients for the series in Fig. \ref{fig:real-data1}. We can see that even with a relatively high order $\color{black}r=9$, the trend functions missed a few peaks and troughs in the data. In other words, there appears to be over smoothing.

\FloatBarrier

\pagebreak
An alternative to fitting polynomial trends is to use B-splines and the results can often get better. A practitioner will face a choice of trend models, while remembering not to ignore dependence in the errors, especially when they have prediction in mind. {\color{black} There is also the problem of overfitting to avoid, which is for the fitted curves to follow the data too closely.}

Recent advances in machine learning have made available to the statistics community a wealth of network structures and the associated training methodologies for finding patterns in vast quantities of data. 
There have been attempts 
at deep learning based statistical forecasting, see \cite{wang2019deep} \cite{salinas2020deepar}\cite{rangapuram2018deep}. 
All these methods require the time series to be independent {\color{black} of each other} so that the loss function can be
written in a simple additive form, thus leaving out dependence information across the series.

In this paper, we model the mean $\bm{\mu}_t$ by a recurrent neural network of the LSTM (Long Short-Term Memory) type, 
and simultaneously $\{{\mathbf y}_t-\bm{\mu}_t\}$ by the VAR model (1). 
All the model parameters are estimated at the same time. The exact Gaussian log-likelihood is used, and no assumption is made on the independence between the component series. We enforce the causality condition on the VAR parameters to ensure the stability of the model, which is often overlooked in the literature. {\color{black}It helps with convergence of the training algorithm and the best linear predictor will be readily available.}

The rest of the paper is organised as follows: 
Section~\ref{trend-generation-VAR-parametrization-using-LSTM} defines the model and discusses trend generation, 
VAR parameterisation, the Gaussian log-likelihood function, and its use in network training. 
Section~\ref{simulation-study-varwt} is a simulation study using trends {\color{black}estimated}
from real data.
Section~\ref{application-study-varwt} shows results of model fitting to three data sets and comparisons with alternative models in terms of forecasting accuracy.
Section~\ref{conclusion-varwt} offers concluding remarks.

\section{Model fitting and prediction}
\label{state-space}

The Deep VAR with Trend (DeepVARwT) model is given by 
 \begin{equation}
 \label{varwt}
 \begin{aligned}
\mathbf{y}_{t}- \bm{\mu}_{t}=A_{1} (\mathbf{y}_{t-1}-\bm{\mu}_{t-1})+A_{2} (\mathbf{y}_{t-2}-\bm{\mu}_{t-2})+\cdots+A_{p} (\mathbf{y}_{t-p}-\bm{\mu}_{t-p})+\bm{\varepsilon}_{t}, \qquad   t=0, \pm1, \pm2.   \ldots, 
\end{aligned} 
\end{equation}
where $\{\bm{\varepsilon}_{t}\}$ is i.i.d. Gaussian vector white noise  with mean vector $\bf 0$ and variance-covariance matrix $\Sigma$. 
It is also assumed that {\color{black}as a result of causality,}   
$\bm {\varepsilon}_t$ is uncorrelated with ${\mathbf y}_{t-1}$, 
${\mathbf y}_{t-2}$,\ldots, so that the RHS of~(\ref{varwt}) consists of the best linear predictor $\hat{\textbf{y}}_{t}- \bm{\mu}_{t}$ of  $\textbf{y}_{t}- \bm{\mu}_{t}$ in terms of $\textbf{y}_{t-1}$, $\textbf{y}_{t-2}$,\ldots (infinite past) and the prediction error $\bm{\varepsilon}_{t}$. 
The trend $\bm{\mu}_{t}$ as well as $A_{1},...,A_{p}$ and 
$\Sigma$ will all come from an LSTM network which is described below. 
{\color{black}In particular, $\bm{\mu}_t$ is to be given in (\ref{mu_t}).}

The difference between this model and the VARwT model (\ref{varx}) is in the formulation of 
$\bm{\mu}_t$. If we leave $\bm{\mu}_t$ unspecified, we have a semi-parametric model.

\label{trend-generation-VAR-parametrization-using-LSTM}
\subsection{Long Short-Term Memory (LSTM)}
\label{lstm}
A neural network takes input $\mathbf{x}_{t}$ at time $t$, passes it through layers of neurons (processing units) to produce an output. A weighted average of all the input received at each neuron goes into an activation function to produce output for the next stage. The combined effect is a nonlinear relationship between the input and the output.

\begin{figure}[h!]
	\centering
	\includegraphics[width=0.6\linewidth]{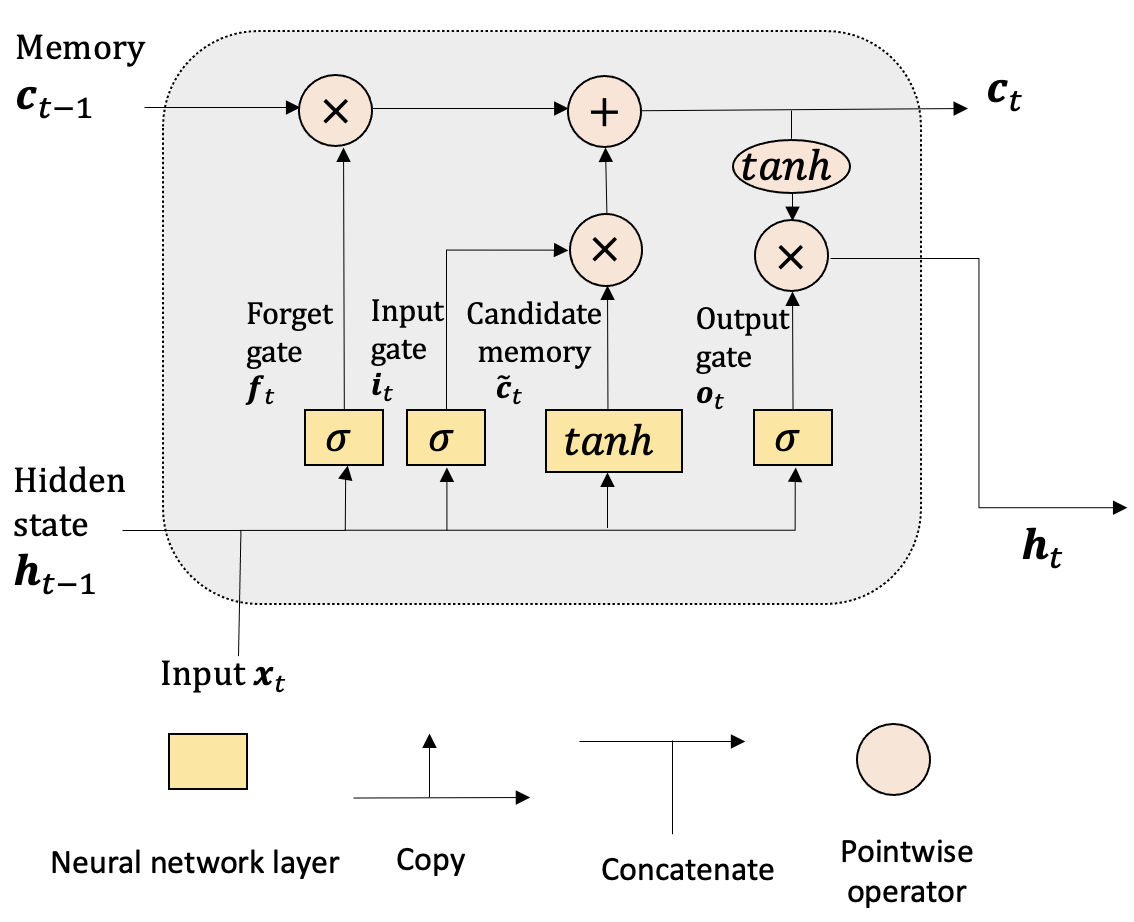}
	\caption{The computation unit for hidden state $\bm{h}_{t}$ in an LSTM.}
	\label{hidden-states}
\end{figure}

{\color{black}A recurrent network also uses the output at time $t-1$ as additional input for time $t$. To overcome the vanishing gradient problem \citep{graves2008novel}, a long short-term memory (LSTM) network uses special cells and gates to control information flow \citep{hochreiter1997long, gers2000learning}}. At time $t$,
the memory cell $\mathbf{c}_{t} $ puts information from the last memory cell $\mathbf{c}_{t-1}$ through the forget gate $\mathbf{f}_{t}$ and information from the candidate memory cell $\tilde{\mathbf{c}}_{t}$ through the input gate $\mathbf{i}_{t}$. The output gate $\mathbf{o}_{t} $ decides how much information from the memory cell $\mathbf{c}_{t} $ should contribute to the hidden state $\mathbf{h}_{t}$. Fig. \ref{hidden-states} shows the computation unit for the hidden state $\mathbf{h}_{t}$ in an LSTM network and the corresponding calculations are as follows \citep{gers2000learning}:
\begin{equation}
\begin{aligned}
\textbf{Input gate:} \quad \quad \mathbf{i}_{t} &=\sigma\left({W}_{x i}\mathbf{x}_{t}+{W}_{h i}\mathbf{h}_{t-1}+\mathbf{b}_{i}\right), \\
\textbf{Forget gate:} \quad \quad \mathbf{f}_{t} &=\sigma\left({W}_{x f}\mathbf{x}_{t} +{W}_{h f}\mathbf{h}_{t-1} +\mathbf{b}_{f}\right), \\
\textbf{Output gate:}\quad \quad \mathbf{o}_{t} &=\sigma\left({W}_{x o}\mathbf{x}_{t} +{W}_{h o}\mathbf{h}_{t-1} +\mathbf{b}_{o}\right),\\
\textbf{Candidate memory cell:} \quad \quad  \tilde{\mathbf{c}}_{t}&=\tanh \left({W}_{x c}\mathbf{x}_{t} + {W}_{h c}\mathbf{h}_{t-1}+\mathbf{b}_{c}\right),\\
\textbf{Memory cell:} \quad \quad \mathbf{c}_{t} &=\mathbf{f}_{t} \odot \mathbf{c}_{t-1}+\mathbf{i}_{t} \odot \tilde{\mathbf{c}}_{t},\\
\textbf{Hidden state:} \quad \quad \mathbf{h}_{t} &=\mathbf{o}_{t} \odot \tanh \left(\mathbf{c}_{t}\right),
\end{aligned}
\label{gates}
\end{equation}
where {\color{black}$\mathbf{x}_{t}$ is the input at time $t$}, ${W}_{x i}$, ${W}_{x f}$, ${W}_{x o}$, ${W}_{h i}$ , ${W}_{h f}$, ${W}_{h o}$, ${W}_{x c}$ and ${W}_{h c}$ are weight parameters, $\mathbf{b}_{i}$, $\mathbf{b}_{f}$, $\mathbf{b}_{o}$ and $\mathbf{b}_{c}$ are bias parameters,  $\sigma(\cdot)$ is the sigmoid function and the operator $\odot$ denotes the element-wise product. {\color{black}For simplicity, we can write 
\begin{equation}
\mathbf{h}_{t}=\mbox{LSTM}\left(\mathbf{h}_{t-1}, \mathbf{x}_{t}; \bm{\Phi}\right), 
\label{h_t}
\end{equation}
where $\bm{\Phi}$ contains all the weight and bias parameters in (\ref{gates}). }
\FloatBarrier

\subsection{Time-dependent trend generation using LSTM}
We map the hidden state $\mathbf{h}_{t}$ to the trend term  $\bm{\mu}_{t}$ as follows: 
\begin{equation}
	\begin{aligned}
		\bm{\mu}_{t}=W_{\mu} \mathbf{h}_{t}+ \mathbf{b}_{\mu},
	\end{aligned}
	\label{mu_t}
\end{equation}
where $W_{\mu}$ and $\mathbf{b}_{\mu}$ are additional weight and bias parameters respectively. {\color{black}Typically $\mathbf{x}_{t}=(t, t^2,\ldots,t^k)'$ with $t$ scaled by the series length $T$ when $k$ is large. Negative powers of $t$ can be included and other basis functions can be used instead.}  
\pagebreak
\subsection{VAR parameter generation}
Let $m$ be the dimension of $\mathbf{y}_t$ and $p$ the order of the VAR model for $\{\mathbf{y}_t-\bm{\mu}_t\}$. We allocate $m^{2}p$ {\color{black} additional} parameters in the neural network to form {\color{black} $m\times m$ matrices ${\mathcal A}_1, \ldots, {\mathcal A}_p$}, and another $m(m+1)/2$ parameters to form a lower triangular matrix $L$ to construct $\Sigma=LL^{\prime}$ for the variance-covariance matrix of $\bm{\varepsilon}_t$. 
The total number of VAR parameters is thus $m^{2}p+m(m+1)/2$.
These parameters 
are initialised and updated by the network together with other network parameters. {\color{black}The matrices ${\mathcal A}_1, \ldots, {\mathcal A}_p$} go through the next step of reparameterisation to become ${A}_{1},...,{A}_{p}$.

\subsection{Reparameterising VAR($p$) to enforce causality}
\label{Reparameterising-algorithm}

It is usually assumed that model~(\ref{var}) is {\color{black}`causal'} in the sense that {\color{black}a stationary solution $\{\mathbf{y}_t\}$ exits and} can be expressed linearly in terms of $\bm{\varepsilon}_{t}$, $\bm{\varepsilon}_{t-1}$,..., so that {\color{black} $\bm{\varepsilon}_{t}$ is uncorrelated with $\mathbf{y}_{t-1}$, $\mathbf{y}_{t-2}$, \ldots. This implies that }
  $\bm{\varepsilon}_t$ is the innovation or one-step-ahead prediction error {\color{black} and the best linear predictor of ${\mathbf y}_t$ is}  
 $\hat{\mathbf y}_t=A_{1} \mathbf{y}_{t-1}+A_{2} \mathbf{y}_{t-2}+\cdots+A_{p} \mathbf{y}_{t-p}$ in terms of the infinite past.   
 The condition for a model to be causal is that all the roots of 
 $
\mbox{det}(I-A_1 z-A_2 z^2-\cdots-A_p z^p)
 $ lie outside the unit circle, {\color{black}c.f. \citep{hannan2009multiple} for the proof}. {\color{black} This is also known as the stationarity or stability condition. Any time series satisfying such a model is guaranteed to be asymptotically stationary, and} the linear system given by (\ref{var}) will be  stable in the sense that bounded input leads to bounded output.
 
{\color{black}A causal model is needed for the best linear predictor coefficients.} The parameter space of a causal VAR model is highly complicated. In the univariate case it can be mapped to $(-1,1)$ in each dimension using partial {auto}correlations, see \cite{barndorff1973parametrization}. Work on the multivariate case include \cite{morf1978covariance}, \cite{ansley1986note},   \cite{roy2019constrained} and \cite{heaps2023enforcing}. 

Given a set of {\color{black} $m\times m$ matrices ${\mathcal A}_{1},\ldots,{\mathcal A}_{p}$ from the neural net}, we transform them using the Ansley-Kohn transform \cite{ansley1986note} in the following two steps, so that the causality condition is satisfied. 
\begin{itemize}
	\item  \textbf{Partial autocorrelation matrix construction.} For $j=1, \ldots, p,$  find the Cholesky factorisation { $I+{\color{black}\mathcal A}_{j} {\color{black}\mathcal A}_{j}^{\prime}=B_{j}B_{j}^{\prime}$, then compute}
	\begin{equation}
	\label{from-A-to-P}
	\begin{aligned}
{	P_{j}=B_{j}^{-1} {\color{black}\mathcal A}_{j}}
	\end{aligned}
\end{equation}	
	as partial autocorrelation matrices \citep{ansley1986note}.
	\item  \textbf{Causal VAR coefficient generation.}  The partial autocorrelation matrices $\{P_{j}\}$ are mapped into the coefficient matrices $\{A_{si}\}$ and $\{A_{si}^{*}\}$ for forward and backward predictions using $s$ past/future values, with prediction error variance-covariance matrices $\Sigma_{s}$ and $\Sigma_{s}^{*}$ respectively. The answers for $s=p$ are used to calculate ${A}_{1},...,{A}_{p}$.

\textbf{Initialisation}: Make $\Sigma_{0}=\Sigma_{0}^{*}=I,$ and $L_{0}=L_{0}^{*}=I$, $I$ being the identity matrix. {\color{black}This avoids the problem of not knowing the covariance matrix at lag $0$.}

\textbf{Recursion}: For {\color{black}$s=1, \ldots, p$, }
\begin{itemize}
 \tightlist
\item[$\bullet$] Compute 
\begin{equation}
\label{cal-A_ss}
A_{\color{black}s, s}=L_{\color{black}s-1} P_{\color{black}s}(L^*_{\color{black}s-1})^{-1}, \quad A^*_{\color{black}s, s}=L^*_{\color{black}s-1} P_{\color{black}s}^{\prime} (L_{\color{black}s-1})^{-1}.
\end{equation}
\item[$\bullet$] For $i=1, \ldots, \color{black}s-1$ ($\color{black}s>1$), compute
 \begin{equation}
 \label{cal-A_si}
A_{\color{black}s, i}=A_{{\color{black}s-1},i}-A_{\color{black}s, s} A^*_{\color{black}s-1, s-i},\quad 
A_{{\color{black}s}, i}^{*}=A_{{\color{black}s-1},i}^{*}-A^*_{\color{black}s, s} A_{\color{black}s-1, s-i}.
\end{equation}

\item[$\bullet$]  Compute
\begin{equation}
\label{cal-sigma}
\Sigma_{\color{black}s}=\Sigma_{\color{black}s-1}-A_{\color{black}s, s} \Sigma^*_{\color{black}s-1} (A_{\color{black}s, s})^{\prime},
\end{equation}
\begin{equation}
\label{cal-sigma-star}
\Sigma^{*}_{\color{black}s}=\Sigma^{*}_{{\color{black}s-1}}-A^{*}_{\color{black}s, s} \Sigma_{\color{black}s-1}(A^{*}_{\color{black}s, s})^{\prime},
\end{equation}
and obtain their Cholesky factorisations $\color{black}L_{s} L_{s}^{\prime}$ and $\color{black}L_{s}^{*} (L_{s}^{*})^{\prime}$ respectively.
\end{itemize}

\textbf{Causal VAR coefficients}: Compute {\color{black}$T=L L_{p}^{-1}$ as required in Lemma 2.3 of \cite{ansley1986note}, where $L$ is the Cholesky factor of $\Sigma$ (also from the neural net),  and then} 
\begin{equation}
\label{cal-A}
A_{i}={\color{black}T}{A}_{pi}{\color{black}T}^{-1}, \quad i=1, \ldots, p
\end{equation}
{for use as casual VAR($p$) coefficient matrices.}

\end{itemize}

\subsection{The Gaussian log-likelihood}
\label{log-likelihood}
Given that the time series $\{\mathbf{y}_{t}\}$ has a Gaussian structure and AR($p$) dependence, the likelihood of $\mathbf{y}=(\mathbf{y}_{1}^{\prime},...,\mathbf{y}_{T}^{\prime})^{\prime}$ can be written as
\begin{equation}
	\label{loss-function}
\begin{aligned}
 L\left(\bm{\Theta};\textbf{y} \right)=f(\textbf{y}_{1},...,\textbf{y}_{p}) \prod_{t=p+1}^{T}f(\textbf{y}_{t} {\vert} \textbf{y}_{t-1},...,\textbf{y}_{t-p}),
\end{aligned}
\end{equation}
using the joint normal density $f(\mathbf{y}_{1},...,\mathbf{y}_{p})$ and the conditional Gaussian densities $f(\mathbf{y}_{t}{\vert}\mathbf{y}_{t-1}$,..., $\textbf{y}_{t-p})$, $t=p+1,...,T$,
where $\bm{\Theta}$ consists of the parameters $\bm{{\color{black}\Theta}}_{1}=\{\bm{\Phi}, W_{\mu}, \bm{b}_{\mu}\}$ for trend generation and 
$\bm{{\color{black}\Theta}}_{2}$
for the VAR coefficient matrices $A_{1},...,A_{p}$ and the $\Sigma$ matrix. The log-likelihood is 
\begin{equation}
	\label{loss-function}
\begin{aligned}
 \ell(\bm{\Theta} ; \mathbf{y})=-\frac{1}{2}\bigg[n \log (2 \pi)+
 \log {\vert}R_p{\vert}+(\mathbf{y}_{1: p}-\bm{\mu}_{1: p})^{\prime} R_p^{-1}(\mathbf{y}_{1: p}-\bm{\mu}_{1: p})+(T-p) \log {\vert}\Sigma{\vert}+{\textstyle\sum_{t=p+1}^T} \bm{\varepsilon}_t^{\prime} \Sigma^{-1} {\bm\varepsilon}_t \bigg],
 \end{aligned}
\end{equation}
where $n=mT$ ($m$ being the dimension of $\mathbf{y}_t$), 
$R_{p}$ is the variance-covariance matrix of $\mathbf{y}_{1:p}=(\textbf{y}_{1}^{\prime},...,\textbf{y}_{p}^{\prime})^{\prime}$ obtained using standard results \citep{lutkepohl2005new}, $\bm{\mu}_{\color{black}1},...,\bm{\mu}_{T}$ from the output of the neural network and the VAR coefficient matrices $A_{1},...,A_{p}$ are used for the recursive calculation of $\bm{\varepsilon}_{p+1},...,\bm{\varepsilon}_{T}$ according to (\ref{varwt}).

\subsection{Network training}
{\color{black} The basic idea of neural network training is gradient descent, that is to move in the opposite direction of the gradient of the loss function. Instead of the usual sum of squares of errors, we define the loss function as $-\ell(\boldsymbol{\Theta} ; \mathbf{y})$, where $\ell(\boldsymbol{\Theta} ; \mathbf{y})$ is the Gaussian log-likelihood given in (\ref{loss-function}). We use different learning rates 
$\eta_1$ and $\eta_2$ for the different parts of $\bm{\Theta}=(\bm{{\color{black}\Theta}}_1,\bm{{\color{black}\Theta}}_2)$, and modify them at each iteration. The trend parameters are fine-tuned with a smaller learning rate so that the trend terms get updated in small steps to avoid large changes that affect the estimation of the VAR parameters. The algorithm is known as AdaGrad (Adaptive Gradient)~\citep{duchi2011adaptive}, which is available in the 
\proglang{PyTorch} package \citep{NEURIPS2019_9015}. 

\pagebreak
At iteration $k=1,\ldots,K$, the calculation to update $\bm{\Theta}^{(k)}=({\bm{{\color{black}\Theta}}}_{1}^{(k)},{\bm{{\color{black}\Theta}}}_{2}^{(k)})$ takes the following steps, where ${\bm{{\color{black}\Theta}}}_{1}^{(k)}$ consists of ${\bm{\Phi}}^{(k)}$, $W_{\mu}^{(k)}$ and $\mathbf{b}_{\mu}^{(k)}$, and $\bm{{\color{black}\Theta}}_{2}^{(k)}$ holds the elements of ${\color{black}\mathcal A}_{1}^{(k)}, \dots,  {\color{black}\mathcal A}_{p}^{(k)}$ and $L^{(k)}$. 
\begin{itemize}
\item {Compute hidden state ${\mathbf{h}_{t}^{(k)}}=\mbox{LSTM}({\mathbf{h}_{t-1}^{(k)}}, \ \mathbf{x}_{t}; {\bm{\Phi}}^{(k)})$ for each $t=1,\ldots,T$. }
\item{Compute trend term ${\bm{\mu}_{t}^{(k)}}	={W_{\mu}^{(k)}} {\mathbf{h}_{t}^{(k)}}+ {\mathbf{b}_{\mu}^{(k)}}$ for each $t=1,\ldots,T$.}
\item{Compute $P_{1}^{(k)}, \ldots, P_{p}^{(k)}$ from {${\color{black}\mathcal A}_{1}^{(k)}, \dots,  {\color{black}\mathcal A}_{p}^{(k)}$} using~(\ref{from-A-to-P}).}
	\item{Transform $P_{1}^{(k)}, \ldots, P_{p}^{(k)} $ into $A_{1}^{(k)},...,\ A_{p}^{(k)}$ using~(\ref{cal-A_ss}) to~(\ref{cal-A}).}
	\item{Compute $\Sigma^{(k)}=L^{(k)}{{L}^{(k)}}^{\prime}$.  }
        \item {Evaluate loss function $-{\ell}(\bm{\Theta}^{(k)};\mathbf{y})$ at $\bm{\Theta}^{(k)}$
        using~(\ref{loss-function}).}
\item{Compute gradients of the loss function $$\mathbf{g}_{1}^{(k)} ={{\textstyle\frac{\partial}{\partial \bm{{\color{black}\Theta}}_{1}}} (-{\ell}({\bm{\Theta}};\mathbf{y}))\big{\vert} }_{\bm{\Theta}=\bm{\Theta}^{(k)}}, ~\mathbf{g}_{2}^{(k)} ={{\textstyle\frac{\partial}{\partial \bm{{\color{black}\Theta}}_{2}}} (-{\ell}(\bm{\Theta};\mathbf{y}))\big{\vert} }_{\bm{\Theta}=\bm{\Theta}^{(k)}}.$$}  
\item{Compute $G_{1}=\sum_{\tau=1}^{k} \mathbf{g}_{1}^{(\tau)} {\mathbf{g}_{1}^{(\tau)}}{}^{\prime}$ and $G_{2}=\sum_{\tau=1}^{k} \mathbf{g}_{2}^{(\tau)} {\mathbf{g}_{2}^{(\tau)}}{}^{\prime}$.}
\item{Update the trend parameters 
\begin{equation}\bm{{\color{black}\Theta}}_{1}^{(k+1)}=\bm{{\color{black}\Theta}}_{1}^{(k)}-\eta_{1} {\mbox{diag}(G_{1})}^{-\frac{1}{2}}\odot\mathbf{g}_{1}^{(k)}.
\end{equation}}  
\item{Update the VAR parameters 
\begin{equation}\bm{{\color{black}\Theta}}_{2}^{(k+1)}=\bm{{\color{black}\Theta}}_{2}^{(k)}-\eta_{2} {\mbox{diag}(G_{2})}^{-\frac{1}{2}}\odot\mathbf{g}_{2}^{(k)}.\end{equation} }   
\end{itemize}
The above is repeated until both 
$$\mbox{
$r c_1=\big {\vert}\frac{\ell\left(\boldsymbol{\Theta}^{(k-1)} ; \mathbf{y}\right)-\ell\left(\boldsymbol{\Theta}^{(k-2)} ; \mathbf{y}\right)}{\ell\left(\boldsymbol{\Theta}^{(k-2)} ; \mathbf{y}\right)}\big{\vert}$ and 

$rc_{2}=\big{\vert} \frac{\ell(\bm{\Theta}^{(k)};\ \mathbf{y})-\ell(\bm{\Theta}^{(k-1)};\ \bm{y} )}{\ell(\bm{\Theta}^{(k-1)};\ \mathbf{y})}\big{\vert}
$}$$
fall below $\mbox{\em prec}$ or iteration $K$ has been reached. 

The initial values $\bm{{\color{black}\Theta}}_1^{\text {(0) }}$ for the trend part are obtained by minimising the sum of squares of differences between $\mathbf{y}_t$ and $\bm{\mu}_t$ , and {$\bm{{\color{black}\Theta}}_{2}^{(0)}=\{{\color{black}\mathcal A}_{1}^{(0)},...,{\color{black}\mathcal A}_{p}^{(0)}, L^{(0)}\}$} by fitting VAR($p$) to the detrended data {using OLS} {\color{black} then applying the inverse Ansley-Kohn transform}. The initial state $\textbf{h}_{0}$ and the initial candidate memory cell $\textbf{c}_{0}$ are both set to $\textbf{0}$. 
}

\subsection{Prediction from trained network}

We continue to run the trained network for $t=T+1$, $T+2$ etc to generate future trend values $\bm{\mu}_{t}$ and produce point forecasts using the formula~(\ref{best-linear-predictor}) in the Appendix. 

Approximate 95\% prediction intervals can be obtained by adding or  subtracting 1.96 times the standard deviations of prediction errors using results in the Appendix.

\pagebreak

\section{Simulation study}
\label{simulation-study-varwt}
To assess the finite sample performance of the deep learning based maximum likelihood estimation method, we simulated 100 samples {\color{black}of 
$\mathbf{y}_{1},...,\mathbf{y}_{T}$ with $T=800$} from the {{semi-parametric VAR(2) model} \color{black}(m=3, p=2)}   
 \begin{equation}
 \label{simu-model}
  \begin{aligned}
\mathbf{y}_{t}- \bm{\mu}_{t}=A_{1} (\mathbf{y}_{t-1}-\bm{\mu}_{t-1})+A_{2} (\mathbf{y}_{t-2}-\bm{\mu}_{t-2})+\bm{\varepsilon}_{t}, \quad 
  \end{aligned} 
\end{equation}
where
\[{A_{1}}=\begin{pmatrix*}[r]-1.0842& -0.1245& 0.3137  \\
-0.7008& -0.3754& -0.2064  \\
0.3166&  0.3251& 0.2135 \\
\end{pmatrix*}, \quad
{A_{2}}=\begin{pmatrix*}[r]-0.5449& -0.3052& -0.1952   \\
-0.4057&  0.5129&  0.3655  \\
0.0054& -0.2911&  0.2066 \\\end{pmatrix*},
\]
and
\[
\Sigma=\begin{pmatrix*}[r]0.4834 & -0.2707  & 0.1368  \\
-0.2707 & 0.4079 & -0.0221  \\
0.1368  & -0.0221  & 0.4103  \\
\end{pmatrix*}
\]
is the variance-covariance matrix of $\bm{\varepsilon}_{t}$. Values of the trend term $\bm{\mu}_{t}$ were obtained by kernel smoothing from daily closing prices of three US stocks from 3rd October 2016 to 5th December 2019.

\begin{figure}[h!]
	\centering
	\includegraphics[width=0.8\linewidth]{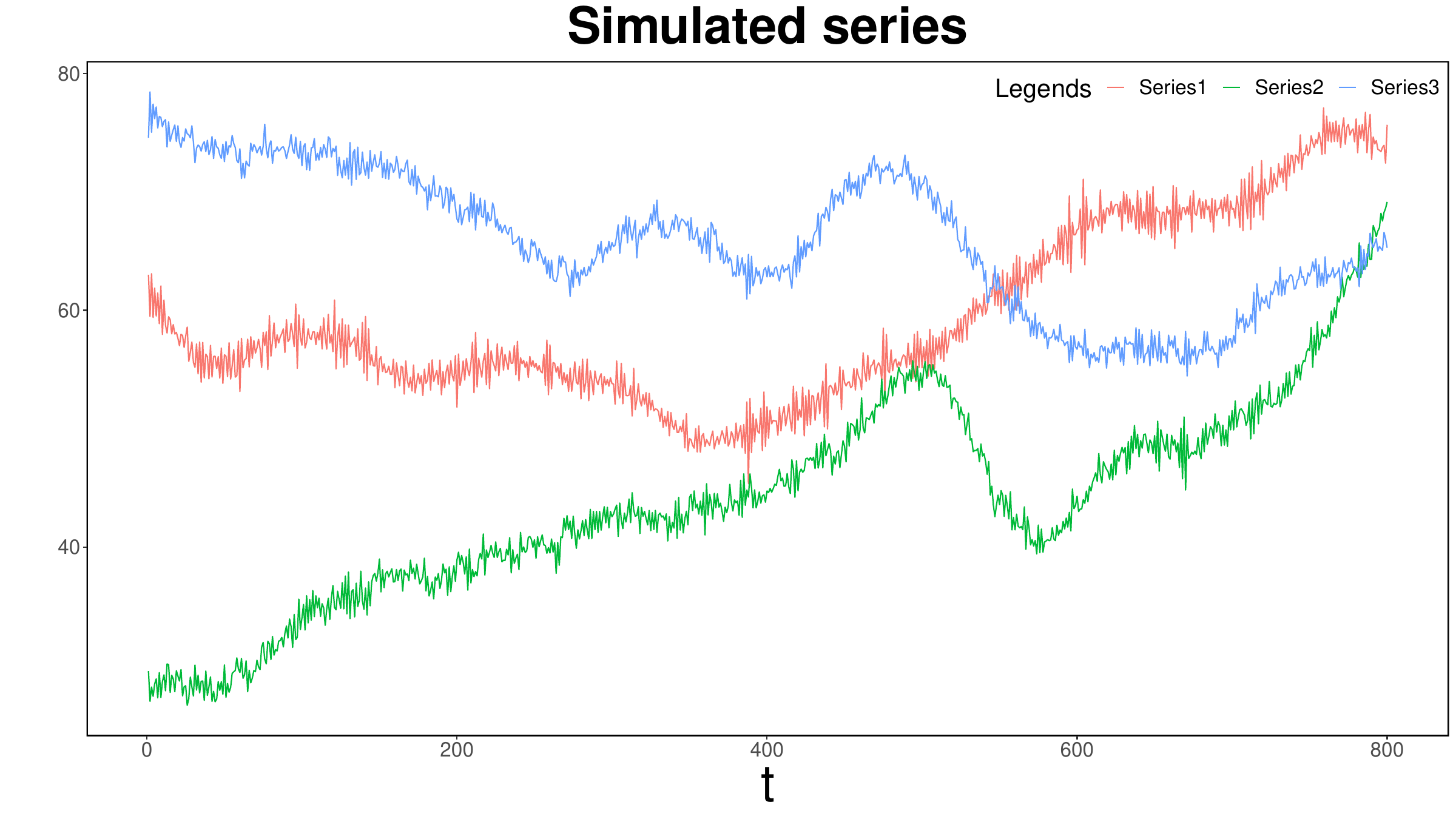}
	\caption{Simulated series from VAR(2) model with trend.} 
	\label{simulated-data-varwT}
\end{figure}

An example of the simulated multiple series is shown in Fig.~\ref{simulated-data-varwT}, each having clearly a trend that looks more realistic than artificial functions. 
\FloatBarrier
To keep the LSTM network simple yet large enough for the trends we used a single hidden layer of 20 units  (Reducing this to $18$ and the results are not as good). The input at time $t$ was $\mathbf{x}_{t}=(t, t^{2}, t^{3},t^{-1}, t^{-2}, t^{-3})'$ (Results are much worse without the negative powers of $t$).
The learning rates were $\eta_{1}=0.001$ and $\eta_{2}=0.01$, with $K=600$ iterations and precision $\mbox{\em prec}=10^{-5}$.  The total number of parameters is $2327$ including $21$ for the VAR($2$).
 The computation time for 100 sets of parameter estimates was about 5.5 hours on an Intel Core i9 2.3 GHz processor with eight cores used in parallel. 

\subsection{Simulation results}
Following \cite{fan2003adaptive}, we use the mean absolute deviation 
\begin{equation*}
   {\mathrm{MAD}_i}=\frac{1}{3\times 800} \sum_{{\color{black}j}=1}^{3} \sum_{t=1}^{800} \big{\vert}\hat{{\mu}}^{(i)}_{{\color{black}{j}}t}-{\mu}_{{\color{black}{j}}t}\big{\vert}
\end{equation*}
to evaluate the accuracy of trend estimation in the $i$th simulation run, where $i=1,...,100$. Fig.~\ref{estimated-trends-from-DeepVARwT-VARwT}  shows estimated trends from DeepVARwT (left panel) and VARwT (right panel) with $\mathrm{MAD}$ at the first quartile (short dashed, black), the median (dotted, red), and the third quartile (long dashed, black) respectively among the 100 simulation runs.
The estimated trends from our model follow the true trends very closely while those from VARwT 
show over-smoothing of local changes. {\color{black}There is no over-fitting here because the trends truly exist {\color{black}in the simulated data.}

\begin{figure}[h!]
	\centering
	\includegraphics[width=0.82\linewidth]{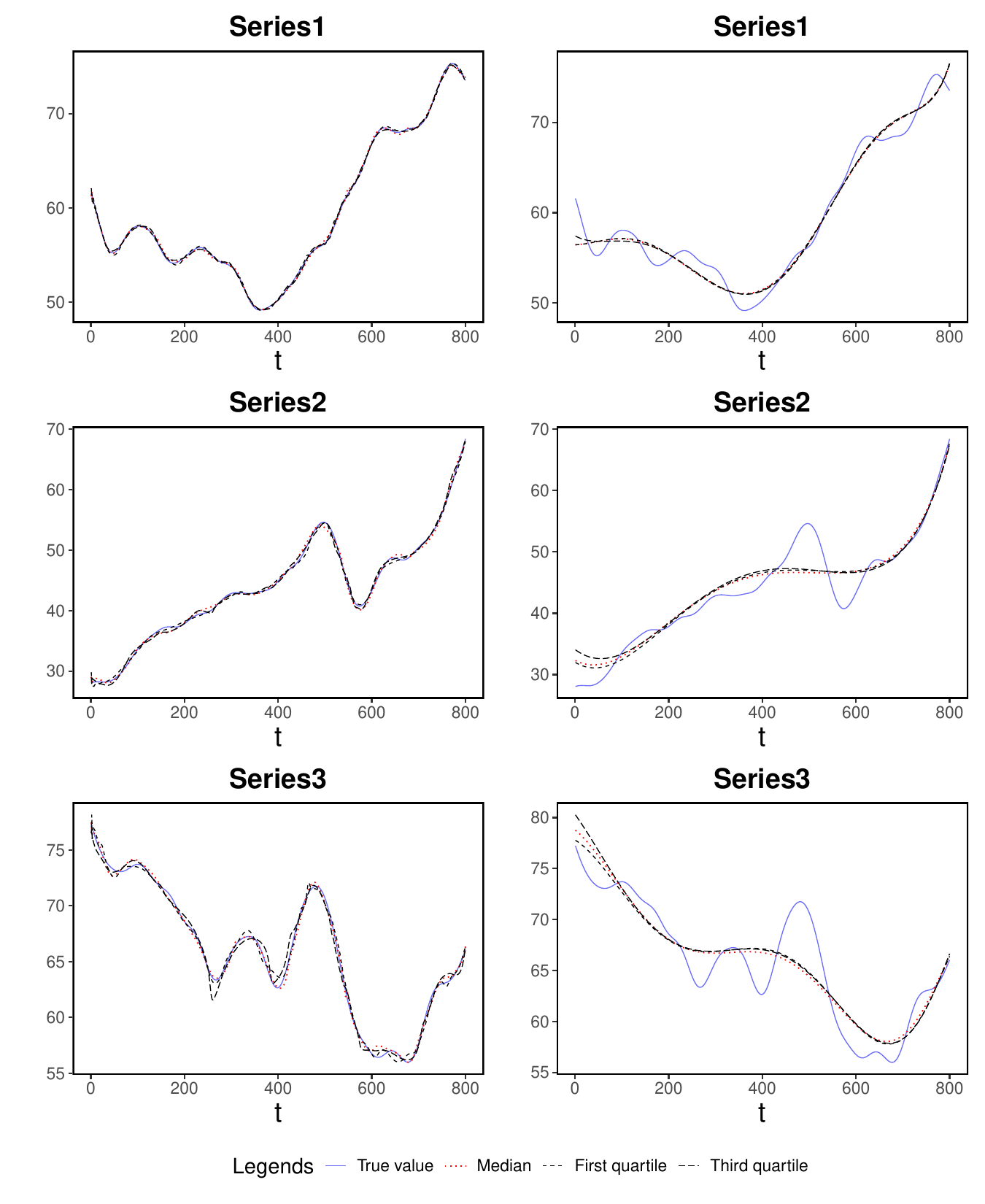}
		\vspace*{-5mm}
	\caption{True (solid, blue) and estimated trends from DeepVARwT (left pane) and VARwT (right pane) with $\mathrm{MAD}$ at first quartile (short dashed, black), third quartile (long dashed, black), and median (dotted, red).}
	\label{estimated-trends-from-DeepVARwT-VARwT}
\end{figure}
\FloatBarrier

Table~\ref{estimated-res} reports summary statistics of 100 estimates of each parameter from our model and VARwT {\color{black}($r=8$)}, where ${{a}^{(i)}_{jk}}$ and $\sigma_{jk}$ refer to the $(j,k)$-th entry of the coefficient matrix $A_{i}$ and the variance-covariance matrix $\Sigma$, respectively. We can observe that compared with VARwT, the DeepVARwT model gives rise to reduced biases at the expense of standard deviations (SDs). The parameter estimates are more accurate with smaller mean squared errors (MSEs) than those obtained from the VARwT model.

\begin{table}[h!]
		\normalsize
		\caption{Estimation results of DeepVARwT and VARwT: true value above sample mean, standard deviation, mean squared error of 100 estimates of each parameter and sample bias.}
		\label{estimated-res}
		\begin{adjustbox}{width=1\textwidth}
		\begin{tabular}{rrrrrrrrrrrrrr}
				\toprule

	&${{a}^{(1)}_{11}}$&${{a}^{(1)}_{12}}$&${{a}^{(1)}_{13}}$&${{a}^{(1)}_{21}}$ &${{a}^{(1)}_{22}}$&${{a}^{(1)}_{23}}$&${{a}^{(1)}_{31}}$&${{a}^{(1)}_{32}}$&${{a}^{(1)}_{33}}$\\
        True value & -1.0842 & -0.1245 & 0.3137 & -0.7008 & -0.3754 & -0.2064 & 0.3166 & 0.3251 & 0.2135 \\ 
        		\cmidrule(lr){1-10}
         &\multicolumn{10}{c} {DeepVARwT}\\
         \cmidrule(lr){1-10}
        Mean &-1.0280&-0.1280&0.2704&-0.6864&-0.3422&-0.2195&0.2923&0.3611&0.2884 \\ 
        Bias &0.0562&-0.0035&-0.0433&0.0144&0.0332&-0.0131&-0.0243&0.0360&0.0749 \\ 
        SD &0.1468&0.0679&0.0734&0.0640&0.0786&0.0530&0.0398&0.0458&0.0974 \\ 
        
        MSE&0.0245&0.0046&0.0072&0.0043&0.0072&0.0030&0.0022&0.0034&0.0150\\
        		\cmidrule(lr){1-10}
         &\multicolumn{10}{c} {VARwT}\\
         \cmidrule(lr){1-10}
        Mean &0.0133& 0.3386 &0.0185&-0.2269&0.1029&-0.4148&0.1871&  0.3508&0.5027  \\ 
          Bias &1.0975&	0.4631	&-0.2952	&0.4739	&0.4783	&-0.2084	&-0.1295	&0.0257	&0.2892  \\ 
        SD &0.0179& 0.0457& 0.0417&0.0178& 0.0266& 0.0239&0.0128& 0.0187& 0.0228\\ 
      
        MSE& 1.2047& 0.2166& 0.0889& 0.2249&0.2295& 0.0440&0.0169& 0.0010&0.0841\\
	\toprule

	&${{a}^{(2)}_{11}}$&${{a}^{(2)}_{12}}$&${{a}^{(2)}_{13}}$&${{a}^{(2)}_{21}}$ &${{a}^{(2)}_{22}}$&${{a}^{(2)}_{23}}$&${{a}^{(2)}_{31}}$&${{a}^{(2)}_{32}}$&${{a}^{(2)}_{33}}$\\
         True value &  -0.5449 & -0.3052 & -0.1952 & -0.4057 & 0.5129 & 0.3655 & 0.0054 & -0.2911 & 0.2066 \\ 
         		        		\cmidrule(lr){1-10}
         &\multicolumn{10}{c} {DeepVARwT}\\
         \cmidrule(lr){1-10}
        Mean &-0.4886&-0.3244&-0.2227&-0.3756&0.5405&0.3741&-0.0229&-0.2648&0.2658 \\
         Bias &0.0563&-0.0192&-0.0275&0.0301&0.0276&0.0086&-0.0283&0.0263&0.0592 \\ 
       SD &0.1698&0.0625&0.0460&0.0840&0.0761&0.0409&0.0524&0.0368&0.0813\\ 
        
         MSE&0.0317&0.0042&0.0029&0.0079&0.0065&0.0017&0.0035&0.0020&0.0100\\

			\cmidrule(lr){1-10}
         &\multicolumn{10}{c} {VARwT}\\
         \cmidrule(lr){1-10}
        Mean &0.7127&-0.2808&-0.1122&0.2335&  0.8189& 0.4848&-0.1662& -0.2880&  0.3966 \\ 
         Bias &1.2576&	0.0244&	0.0830&	0.6392&	0.3060&	0.1193	&-0.1716&	0.0031	&0.1900  \\ 
        SD &0.0206& 0.0468& 0.0435&0.0186& 0.0280& 0.0231&0.0127&0.0177& 0.0233  \\ 
       
        MSE&1.5820& 0.0028& 0.0088&0.4089& 0.0944& 0.0148&0.0296& 0.0003& 0.0366\\
	\toprule
	&$\sigma_{11}$	&$\sigma_{21}$&$\sigma_{22}$&$\sigma_{31}$&$\sigma_{32}$&$\sigma_{33}$\\
	True value&0.4834 &-0.2707&0.4079&0.1368&-0.0221&0.4103\\
	\cmidrule(lr){1-7}
          &\multicolumn{7}{c} {DeepVARwT}\\
         \cmidrule(lr){1-7}
Mean&0.5197&-0.2723&0.4238&0.1194&-0.0282&0.4317 \\
Bias &0.0363&-0.0016&0.0159&-0.0174&-0.0061&0.0214        \\
SD&0.0928&0.0451&0.0436&0.0251&0.0182&0.0369\\

MSE&0.0098&0.0020&0.0021&0.0009&0.0004&0.0018\\
	\cmidrule(lr){1-7}
          &\multicolumn{7}{c} {VARwT}\\
         \cmidrule(lr){1-7}
Mean& 1.2291&  -0.0039&  0.6636& 0.0714&-0.0625&  0.5114 \\
Bias & 0.7457	&0.2668	&0.2557	&-0.0654	&-0.0404	&0.1011            \\
SD& 0.0811&  0.0267& 0.0352&  0.0312& 0.0206& 0.0265     \\

MSE&0.5626&0.0719& 0.0666& 0.0052& 0.0020& 0.0109\\

				\midrule
			\end{tabular}
		\end{adjustbox}

\end{table}
\FloatBarrier

\pagebreak

\section{Real data applications}
\label{application-study-varwt}

\subsection{US macroeconomics series 1}
For the US macroeconomic series  (Fig.~\ref{fig:real-data1}), we fit a model (and make forecasts) 20 times {\color{black}to reduce uncertainty in the accuracy measures}, each time using a {\color{black}time-shifted} training sample of size $T=166$. The training samples are  $y_{1:T}^{(i)}=\{\textbf{y}_{i},\textbf{y}_{i+1},...,\textbf{y}_{i+T-1}\}$, $i=1,...,20$ and we forecast $h=1,2,...,8$ quarters 
ahead. 

The first model we fitted was DeepVARwT(4). The order $p = 4$ for a VAR model is a common choice in the analysis of quarterly macroeconomic series, for example, \cite{koop2010bayesian}, \cite{koop2013forecasting} and \cite{heaps2023enforcing}.
{\color{black}There are hyperparameters to set for the neural net, the most crucial ones being the maximum power of $t$ in the input ${\mathbf x}_t$} and the hidden state size. A grid search was conducted to find 
 the values with maximum likelihood, among {\color{black}powers of} $2$, $3$ or $4$ and $5$, $10$ or $15$ hidden states. For efficiency, we relied on our experience to set values for the other hyperparameters. The learning rates were $\eta_{1}=0.0005$ and $\eta_{2}=0.01$, with $K=500$ iterations and precision $\mbox{\em prec}=10^{-7}$.}
  
\begin{figure}[h!]
	\centering
	\includegraphics[width=0.8\linewidth]{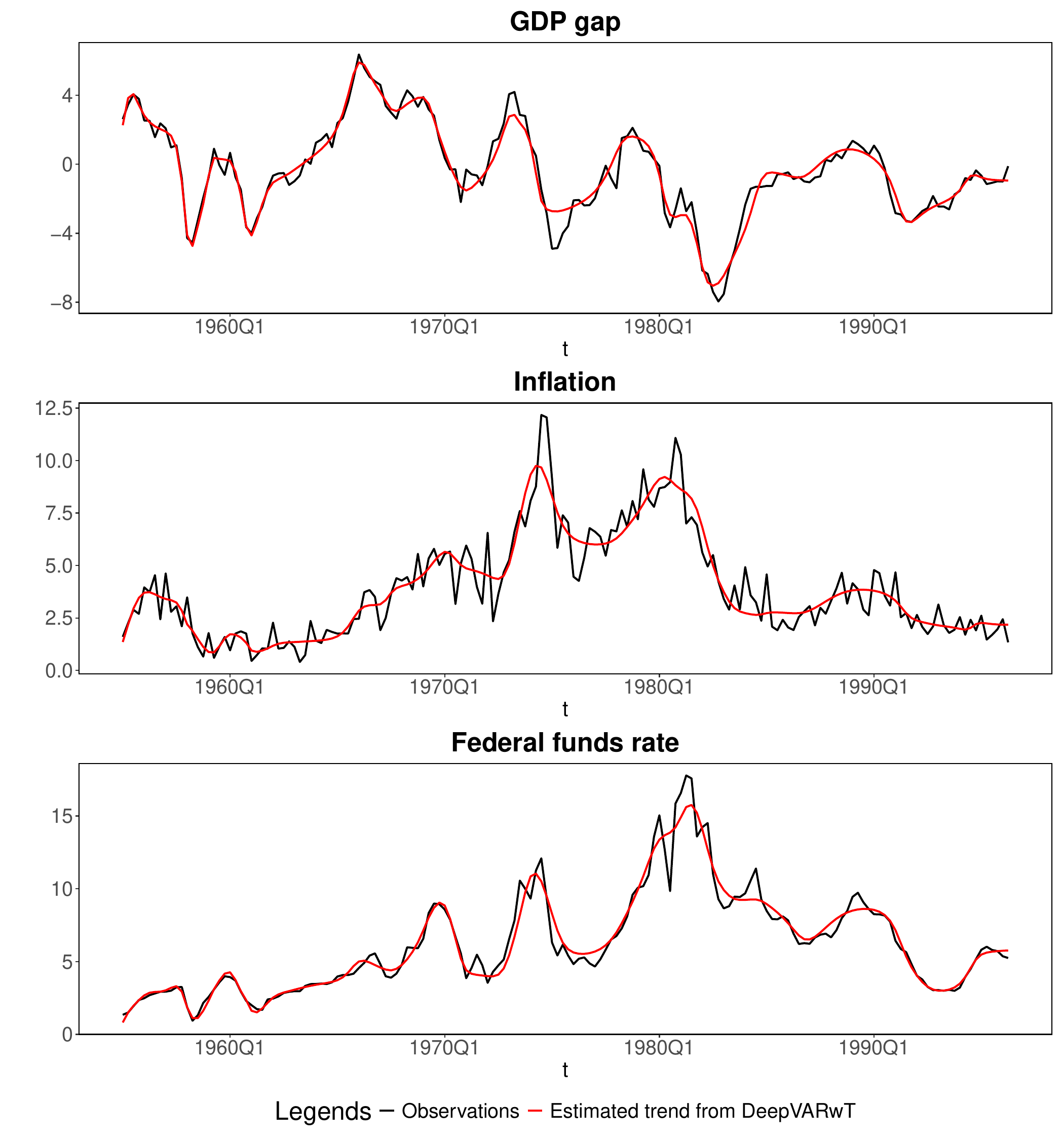}
			\vspace*{-5mm}
	\caption{The first training sample (black lines) from 1955Q1 to 1996Q2 and the corresponding estimated trends (red lines).} 
	\label{fig:estimated-trend-real-data1}
\end{figure}

The estimated trends (red) are shown in Fig.~\ref{fig:estimated-trend-real-data1} for the first training sample $(i=1)$, which can be seen to follow the observations (black) smoothly {\color{black}without overfitting}. 
\FloatBarrier

\begin{figure}[h!]
	\centering
	\includegraphics[width=0.85\linewidth]{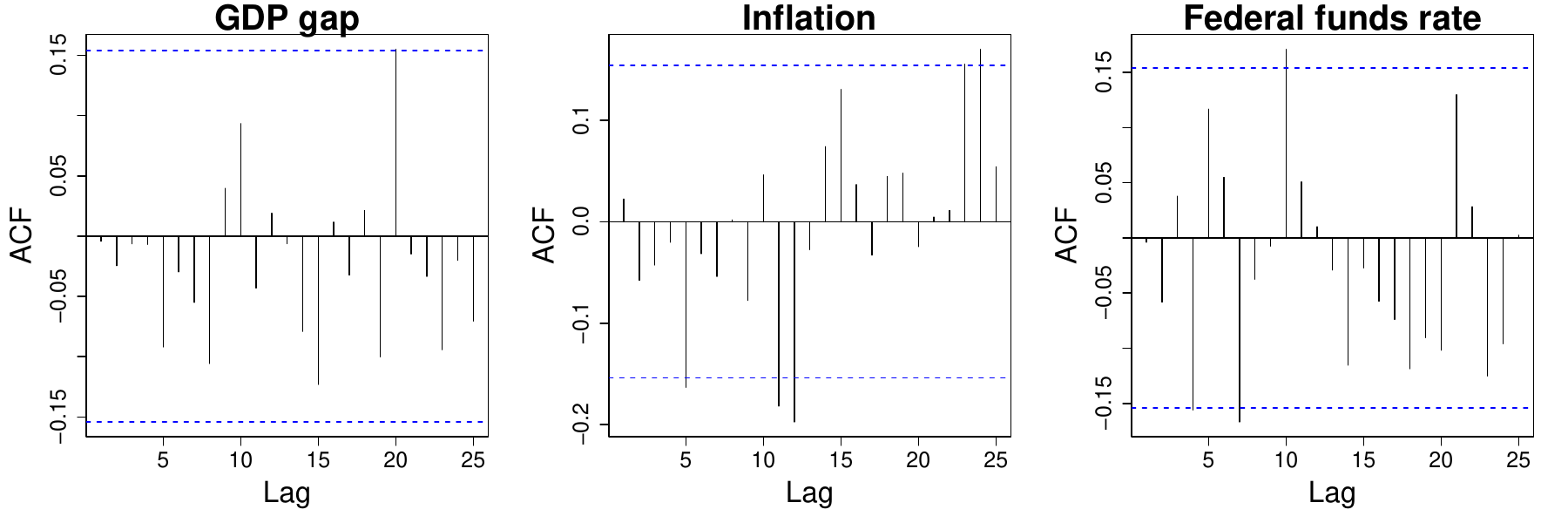}
			\vspace*{-5mm}
	\caption{Sample autocorrelations of residuals.} 
	\label{fig:acf-residuals-section0-us-macro}
\end{figure}

\pagebreak
The sample autocorrelations of residuals are shown in Fig.~\ref{fig:acf-residuals-section0-us-macro}. 
{The results are reasonable for the GDP gap series with only one value out of 25 outside the boundary, and a little concerning for the other two series with one more value each slightly outside the boundaries. }

\FloatBarrier

\begin{figure}[h!]
	\centering
	\includegraphics[width=0.85\linewidth]{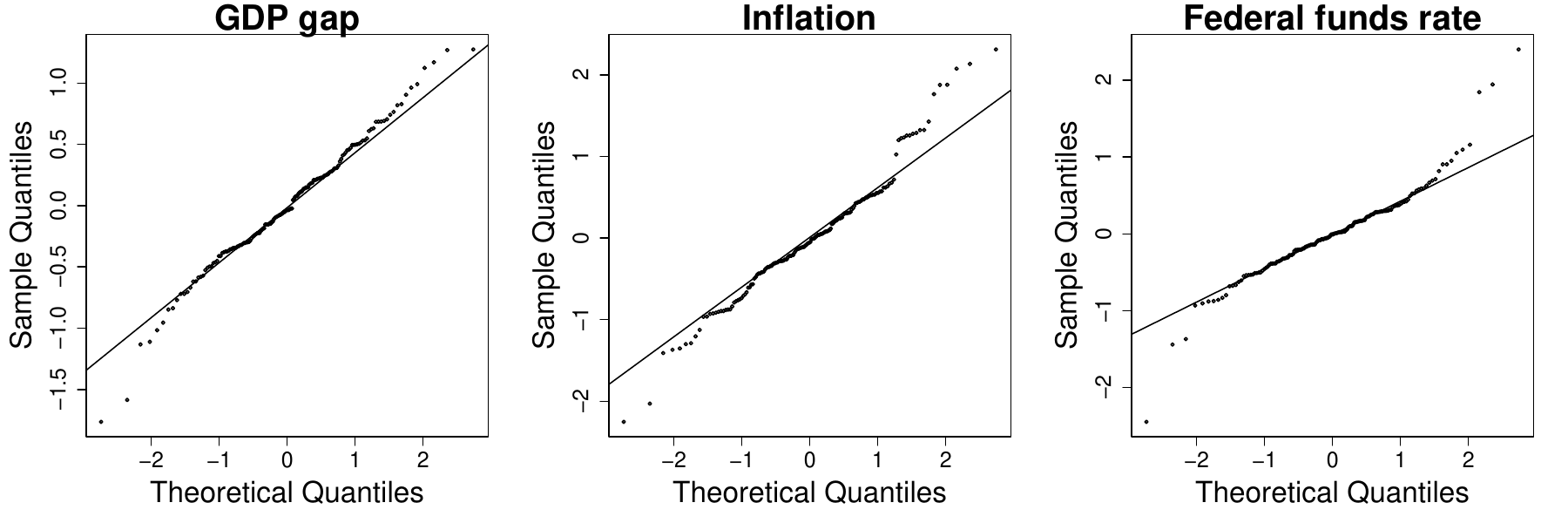}
			\vspace*{-5mm}
 \caption{Normal QQ plots of residuals.}
  \label{fig:qq-plots-us-macro}
\end{figure}

Fig.~\ref{fig:qq-plots-us-macro} contains normal QQ plots of the residuals. There are small deviations from normality for all the series at both ends. 

\FloatBarrier

For comparison, we also fitted a VARwT(4) model, a DeepAR and a DeepState model using default hyperparameter values. {\color{black}The term} $\mathbf{\color{black}z}_{t}=(t, t^{2}, \ldots, t^{9})^{\prime}$ was used in the VARwT model {\color{black}(\ref{varx})} to account for the number of turning points in the trend. 
Table \ref{tab:benchmark-methods} gives a summary of these models and the software packages used.
\begin{table}[ht]
	\caption{Models used for comparison.}
	\label{tab:benchmark-methods}
	\centering
	\resizebox{\textwidth}{!}{
		\begin{tabular}{p{0.23\columnwidth}p{0.5\columnwidth}p{0.38\columnwidth}}
			\toprule
			Model& Description  &Available software           \\
			\midrule
		VARwT               &Vector autoregressive model with trend \citep{vars}  &\code{vars::VAR(exogen=x)}               \\
			DeepAR                &Deep learning based autoregressive model \citep{salinas2020deepar}                                                                                 &\code{gluonts.DeepAREstimator()}                 \\
			DeepState    &Deep learning based state space model \citep{rangapuram2018deep}                           &\code{gluonts.DeepStateEstimator()}                          \\
						\bottomrule
	\end{tabular}}
\end{table}
\FloatBarrier

To evaluate the accuracy of point forecasts, we computed the $h$-step-ahead Absolute Percentage Error averaged over 20 forecasts
\begin{equation*}
	\operatorname{APE}(h)=\frac{1}{20}\sum_{i=1}^{20}\bigg{\vert}\textstyle\frac{y_{T+h}^{(i)}-\hat{y}_{T+h}^{(i)}}{y_{T+h}^{(i)}}\bigg{\vert}\times 100
\end{equation*}
for each component series $\{y_t^{(i)}\}$ in the $i$th training sample, where $\hat{y}_{T+h}^{(i)}$ is the $h$-step-ahead  forecast of ${y}_{T+h}^{(i)}$. The Scaled Interval Score~\citep{Gneiting2007a}
is averaged as follows:
\begin{equation*}
\mathrm{SIS}(h)=\frac{1}{20} \sum_{i=1}^{20} \frac{u_{T+h}^{(i)}-l_{T+h}^{(i)}+\frac{2}{\alpha}(l_{T+h}^{(i)}-y_{T+h}^{(i)}) \mathbbm{1}_{\{y_{T+h}^{(i)}<l_{T+h}^{(i)}\}}+\frac{2}{\alpha}(y_{T+h}^{(i)}-u_{T+h}^{(i)}) \mathbbm{1}_{\{y_{T+h}^{(i)}>u_{T+h}^{(i)}\}}}{\frac{1}{T-s} \sum_{t=s+1}^T {\vert}y_t^{(i)}-y_{t-s}^{(i)}{\vert}},
\end{equation*}
to measure the overall accuracy of the $(1-\alpha) \times 100 \%$ prediction intervals ($l_{T+h}^{(i)}, u_{T+h}^{(i)}$) for the $i$th training sample, $i=1,...,20$, where $\mathbbm{1}_{A}$ is the indicator function for the condition $A$, $s$ is the seasonality of the time series ($s=4$ for quarterly data).

Table \ref{mean-accuracy-uc-macro} shows the forecasting performances of different models at several horizons $h=1,2,4,8$ and averages over $h=1,...,4$ and $h=1,...,8$. When a model performs best, the corresponding number in the table will be bold. 
\begin{itemize}
	\tightlist

		\item \textbf{DeepVARwT vs VARwT.} Compared with VARwT, DeepVARwT produced superior point forecasts at almost all the forecasting horizons for all the series (except $h=1$ for GDP gap and $h=1,2$ for inflation).  Our model also gave more accurate prediction intervals in the long term ($h = 4, 8$) and overall ($h=1:4$ and $h = 1 : 6$) for all the series.
	\item \textbf{DeepVARwT vs other deep learning based models.} Compared with DeepAR and DeepState, our model resulted in better point forecasts and prediction intervals at all the forecasting horizons for federal funds rate.
It also gave more precise prediction intervals at all the forecasting horizons for GDP gap. 

\end{itemize}

	\begin{table}[h]
	\begin{center}
		\normalsize
		\centering
		\caption{Performance of DeepVARwT against other models according to APE and SIS.}
		\label{mean-accuracy-uc-macro}
		\begin{adjustbox}{width=1\textwidth}
			\begin{tabular}{rrrrrrrrrrrrrrrrr}
				\toprule
				&\multicolumn{13}{c} {GDP gap}\\
				
				\cmidrule(lr){2-13} 
				&	\multicolumn{6}{c}{Absolute Percentage Error} &\multicolumn{6}{c}{Scaled Interval Score}  \\
			&$h$=1 &$h$=2&$h$=4&$h$=8&$h$=1:4 &$h$=1:8&$h$=1&$h$=2&$h$=4 &$h$=8&$h$=1:4&$h$=1:8 \\
	\cmidrule(lr){2-7} \cmidrule(lr){8-13} 
        VARwT & 665.927 &2982.609  & 293.199 &1124.228&1042.088&897.853&\textbf{1.592}   &\textbf{4.114}& 34.332 & 244.788 &13.157 &81.907\\ 
        DeepAR&\textbf{279.929}&\textbf{395.924}&170.076&217.204&\textbf{242.408}&\textbf{225.260}&9.438&17.469&36.980&64.358&22.768&39.644\\ 
        DeepState &1023.302&1070.784&178.982&\textbf{200.245}&612.331&403.615&7.903&16.995&22.545&34.583&16.800&24.950 \\ 
        DeepVARwT &684.799&860.708&\textbf{162.429}&202.367&465.475&326.513&4.775&8.864&\textbf{14.426}&\textbf{29.092}&\textbf{10.122}&\textbf{16.942}\\ 
        	\cmidrule(lr){2-7} \cmidrule(lr){8-13} 
       	&\multicolumn{13}{c} {Inflation}\\
				
				\cmidrule(lr){2-13} 
				&	\multicolumn{6}{c}{Absolute Percentage Error} &\multicolumn{6}{c}{Scaled Interval Score}  \\
				&$h$=1 &$h$=2&$h$=4&$h$=8&$h$=1:4 &$h$=1:8&$h$=1&$h$=2&$h$=4 &$h$=8&$h$=1:4&$h$=1:8 \\
	\cmidrule(lr){2-7} \cmidrule(lr){8-13} 
        VARwT &  37.579&  55.549 &  111.768 &348.255&69.868&159.859&\textbf{3.259}&   \textbf{6.236}  &  18.767 & 106.686&9.367&37.482 \\ 
        DeepAR &\textbf{28.859}&\textbf{29.674}&\textbf{43.019}&50.007&\textbf{33.454}&\textbf{42.913}&9.158&9.973&10.320&13.189&9.524&10.887 \\ 
        DeepState &75.165&68.626&75.062&59.455&73.611&69.405&4.554&6.423&\textbf{3.940}&5.121&\textbf{5.317}&6.552 \\ 
        DeepVARwT&66.640&67.879&57.424&\textbf{46.913}&62.047&58.582&8.671&8.729&4.934&\textbf{4.559}&6.713&\textbf{6.369} \\ 
	        	\cmidrule(lr){2-7} \cmidrule(lr){8-13} 
	            	&\multicolumn{13}{c} {Federal funds rate}\\
				
				\cmidrule(lr){2-13} 
				&	\multicolumn{6}{c}{Absolute Percentage Error} &\multicolumn{6}{c}{Scaled Interval Score}  \\
				&$h$=1 &$h$=2&$h$=4&$h$=8&$h$=1:4 &$h$=1:8&$h$=1&$h$=2&$h$=4 &$h$=8&$h$=1:4&$h$=1:8 \\
	\cmidrule(lr){2-7} \cmidrule(lr){8-13} 
        VARwT & 10.521&  25.868 &  90.407& 424.838&44.681&157.662&2.226 &  7.145  &  31.160 &  137.790&14.374&53.498 \\ 
        DeepAR&10.817&22.088&56.348&128.127&31.835&66.314&5.706&12.743&30.447&53.647&17.593&31.701\\ 
        DeepState &33.190&36.782&59.985&128.743&44.186&70.323& 6.963&10.260&13.669&22.901&10.863&15.458\\ 
        DeepVARwT &\textbf{7.800}&\textbf{14.007}&\textbf{33.744}&\textbf{80.529}&\textbf{19.638}&\textbf{39.952}&\textbf{1.468}&\textbf{3.920}&\textbf{8.790}&\textbf{17.309}&\textbf{5.241}&\textbf{9.512}\\
				\midrule
			\end{tabular}
		\end{adjustbox}
	\end{center}
	\end{table}
\FloatBarrier

 The DeepVARwT model outperformed the other models for federal funds rate. It gave the best prediction intervals for GDP gap over h=1:4 and h=1:8, while in second place for prediction accuracy. Its performance is similar for inflation with a slight drop to second place in terms of SIS over h=1:4. 

 All the experiments were conducted on an Intel Core i9 2.3 GHz processor with eight cores. The number of weight parameters in the LSTM network for the 20 fitted DeepVARwT models ranged from 715 to 1350. The computation time for 20 sets of predictions from VARwT, DeepAR, DeepState and DeepVARwT was approximately 1, 42, 210, and 55 minutes, respectively.

\subsection{Global temperatures}
\label{temp-sec}
\begin{figure}[H]
	\centering
	\includegraphics[width=0.7\linewidth, height=4.7in]{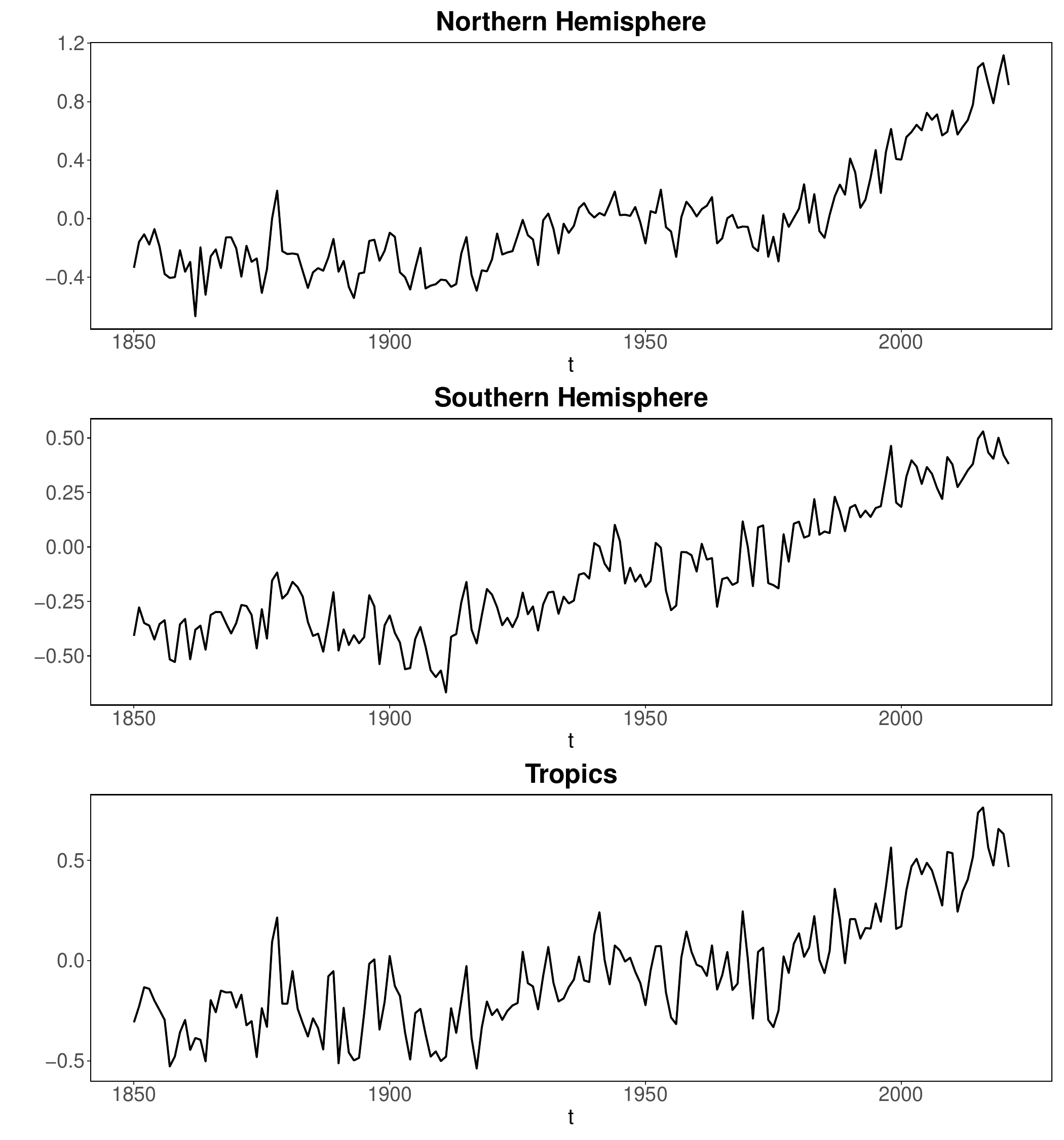}
	\caption{Temperature anomaly series for the Northern Hemisphere, the Southern Hemisphere and the Tropics from 1850 to 2021.} 
	\label{temp}
\end{figure}

Global warming has attracted significant attention in recent research, as demonstrated by studies such as \cite{harvey2001modelling}, \cite{ivanov20101963}, and \cite{holt2020global}. Fig.~\ref{temp} shows three annual temperature anomaly series from distinct regions: the Northern Hemisphere, the Southern Hemisphere and the Tropics from 1850 to 2021, which are described in detail in \cite{morice2012quantifying}. The data are temperature anomalies relative to a reference period of 1961-1990  \citep{morice2012quantifying}. Each series consists of 172 yearly observations.

From Fig.~\ref{temp}, we can observe obvious trends in the three series. \cite{holt2020global} assumed that the trends in the Northern and Southern Hemispheres series are deterministic and modelled the local changes in data using a vector shifting-mean autoregressive model with order $p=3$. We continue to fit a DeepVARwT(3) model to the three series and make predictions $h=1,2,...,6$ steps ahead of $T=147$. As with our first real data application, this is repeated 19 times, each time moving the training sample forward by one time point.
The maximum power of $t$ in ${\mathbf x}_t$ was chosen among $2,3,4$ and the hidden state size among $3,5,8$. The learning rates were $\eta_{1}=0.0005$ and $\eta_{2}=0.01$, with $K=500$ iterations and precision $\mbox{\em prec}=10^{-7}$. 

The forecasts will be compared with those from VARwT(3), DeepAR, 
and DeepState
models with default hyperparameters. The exogenous variables for VARwT are $\mathbf{\color{black}z}_t=(t,t^2,\ldots,t^5)^{\prime}$ to account for the number of turning points in the series. 

\begin{figure}[h!]
	\centering
	\includegraphics[width=0.7\linewidth]{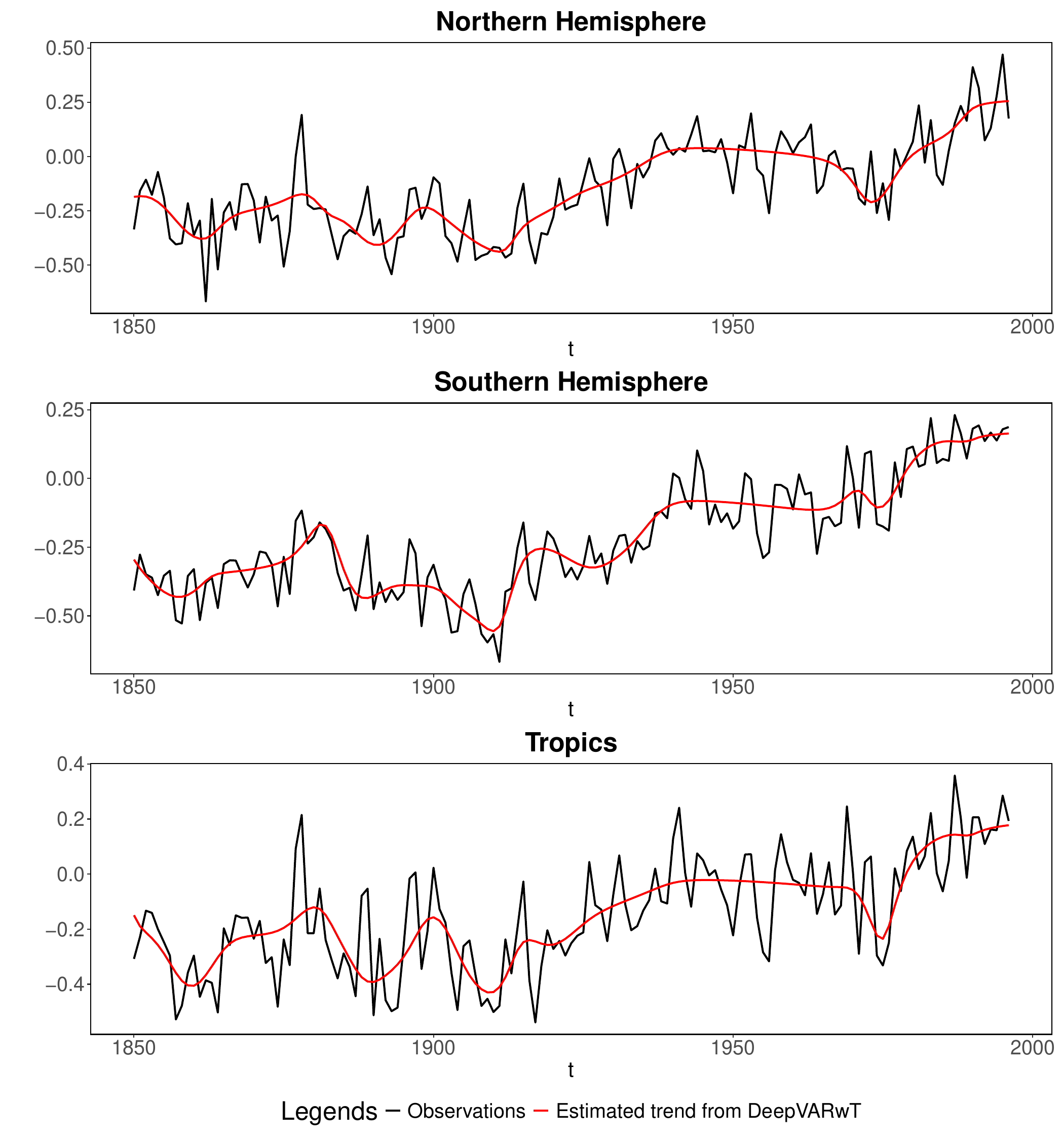}
			\vspace*{-5mm}
	\caption{The first training sample (black lines) from 1850 to 1996 and the corresponding estimated trends (red lines).} 
	\label{fig:estimated-trend-climate}
\end{figure}

From Fig.~\ref{fig:estimated-trend-climate}, we can see that the estimated trends (red) for the first training sample $(i=1)$ follow the observations (black) smoothly.

\FloatBarrier
\begin{figure}[h!]
	\centering
	\includegraphics[width=0.9\linewidth]{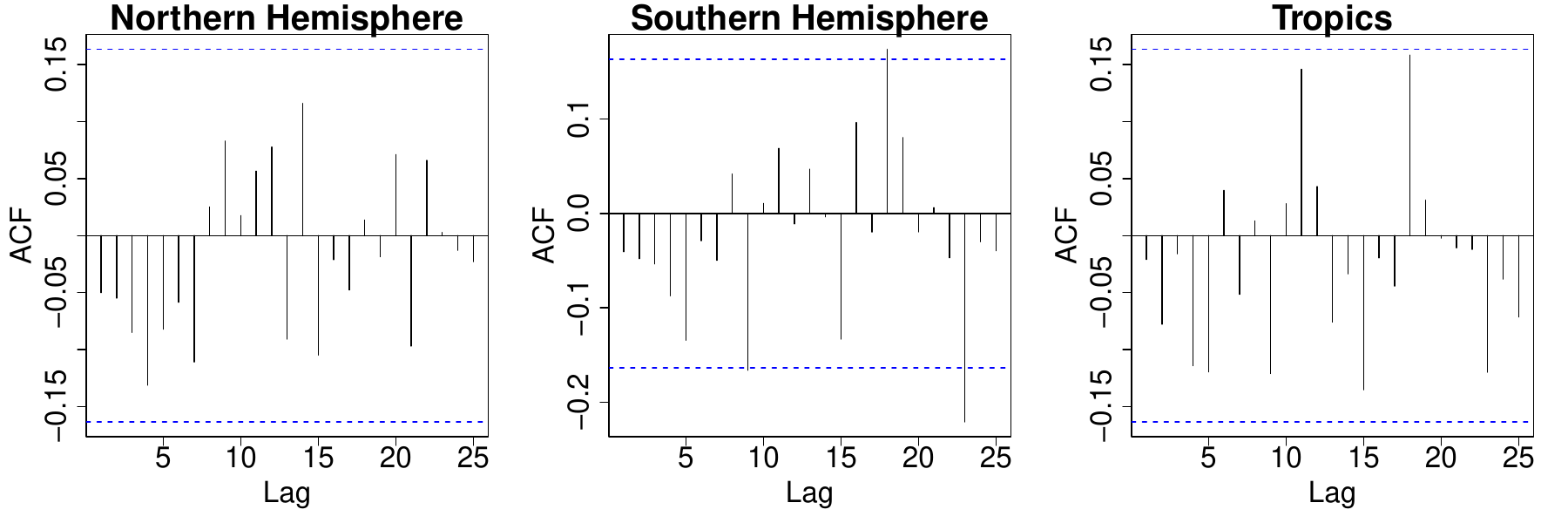}
			\vspace*{-5mm}
	\caption{Sample autocorrelations of residuals.} 
	\label{fig:acf-residuals-section0-climate}
\end{figure}

\FloatBarrier
The sample autocorrelations of residuals are shown in Fig.~\ref{fig:acf-residuals-section0-climate}. 
The results are good for the Northern Hemisphere and the Tropics with all the values within boundaries, and reasonable for the Southern Hemisphere.

\FloatBarrier

\begin{figure}[h!]
	\centering
	\includegraphics[width=0.9\linewidth]{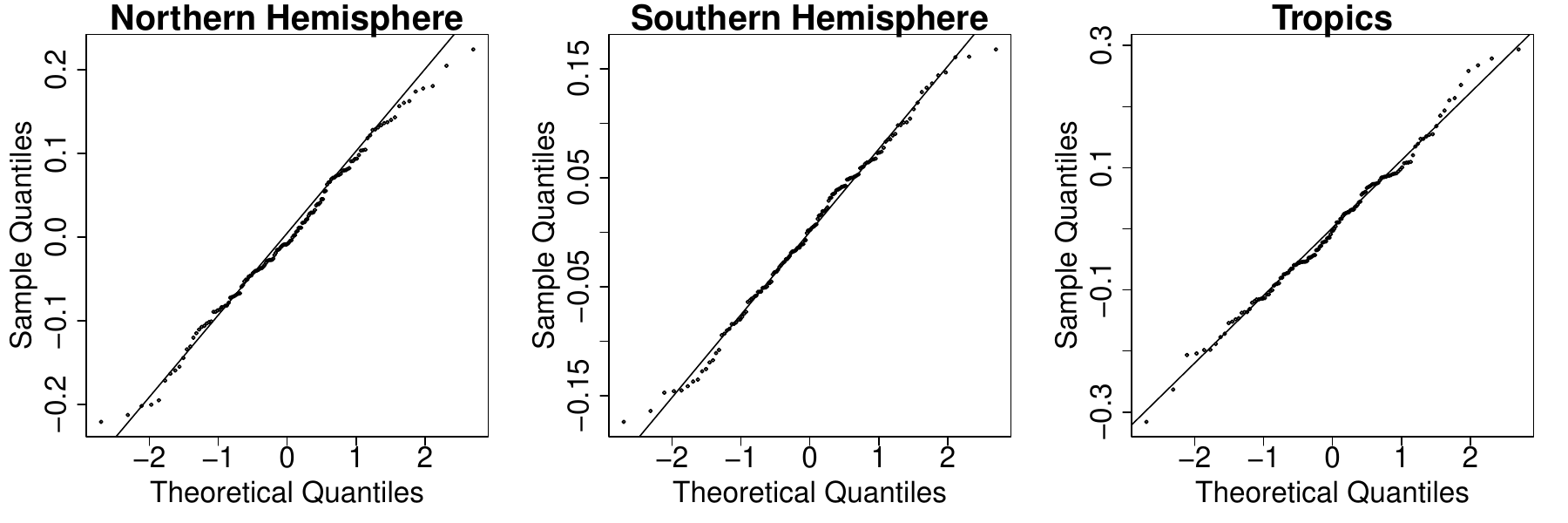}
			\vspace*{-5mm}
 \caption{Normal QQ plots of residuals.}
  \label{fig:qq-plots-climate}
\end{figure}

Fig.~\ref{fig:qq-plots-climate} contains normal QQ plots of the residuals. The results are very good for all the series showing clearly straight line patterns.

\FloatBarrier
Table \ref{mean-accuracy-climate} shows the APE and SIS values of different models at several horizons $h=1,2,4,6$ and averaged over $h=1:3$ and $h=1:6$. 
\begin{itemize}
	\tightlist

		\item \textbf{DeepVARwT vs VARwT.} Compared with the time-invariant VAR with trend, our model produced better point forecasts at all forecasting horizons for all the series.
		It gave better prediction intervals in the long term ($h = 4, 6$) and overall ($h=1:6$) for all the series. 
	\item \textbf{DeepTVARwT vs other deep learning based models.} Compared with DeepAR and DeepState, our model produced more accurate point forecasts  at all forecasting horizons for all the series. Our model resulted in better prediction intervals at all forecasting horizons for all the series.

\end{itemize}

	\begin{table}[H]
	\begin{center}
		\normalsize
		\centering
		\caption{Performance of DeepVARwT against other models according to APE and SIS.}
		\label{mean-accuracy-climate}
		\begin{adjustbox}{width=1\textwidth}
			\begin{tabular}{rrrrrrrrrrrrrrrrr}
				\toprule
    				&\multicolumn{13}{c} {Nothern Hemisphere}\\			
				\cmidrule(lr){2-13} 
				&	\multicolumn{6}{c}{Absolute Percentage Error} &\multicolumn{6}{c}{Scaled Interval Score}  \\
				&$h$=1 &$h$=2&$h$=4&$h$=6&$h$=1:3 &$h$=1:6&$h$=1&$h$=2&$h$=4 &$h$=6&$h$=1:3&$h$=1:6 \\
	\cmidrule(lr){2-7} \cmidrule(lr){8-13} 
  VARwT &26.718& 41.661 &52.713 & 67.507&39.222&49.277&{8.385}& 17.526&  41.703& 75.358&18.991&38.609 \\ 
        DeepAR & 25.485 & 31.243 & 37.544 & 44.097 & 28.563 & 34.805 & 30.180 & 36.423 & 46.395 & 58.302 & 33.410 & 43.010 \\ 
        DeepState & 21.557 & 30.886 & 28.142 & 32.782 & 27.221 & 28.246 & 20.504 & 35.893 & 39.856 & 50.514 & 32.650 & 38.266 \\ 
        DeepVARwT &\textbf{17.261}&\textbf{21.079}& \textbf{19.380}&   \textbf{24.743}&\textbf{19.991}&\textbf{21.250}&\textbf{7.055}& \textbf{17.336}& \textbf{14.320}& \textbf{21.586}&\textbf{14.039}&\textbf{15.471}\\ 
			\cmidrule(lr){2-7} \cmidrule(lr){8-13} 
				&\multicolumn{13}{c} {Southern Hemisphere}\\
				
				\cmidrule(lr){2-13} 
				&	\multicolumn{6}{c}{Absolute Percentage Error} &\multicolumn{6}{c}{Scaled Interval Score}  \\
				&$h$=1 &$h$=2&$h$=4&$h$=6&$h$=1:3 &$h$=1:6&$h$=1&$h$=2&$h$=4 &$h$=6&$h$=1:3&$h$=1:6 \\
	\cmidrule(lr){2-7} \cmidrule(lr){8-13} 
         VARwT &33.082& 47.766&  57.959 & 73.987&45.437&55.317&\textbf{6.252}&  \textbf{9.717}&  14.918&  40.835&{9.313}&18.502 \\ 
        DeepAR & 39.625 & 43.267 & 44.622 & 58.828 & 38.624 & 45.408 & 29.747 & 32.777 & 29.830 & 45.400 & 28.400 & 32.624 \\ 
        DeepState & 28.808 & 43.194 & 26.156 & {36.492} & 35.616 & 33.763 & 25.584 & 40.586 & 24.545 & 36.835 & 33.596 & 32.384 \\ 

        DeepVARwT &\textbf{22.916} &\textbf{25.135}& \textbf{25.006}&  \textbf{34.668}&\textbf{23.030}&\textbf{26.810}&6.470& 11.090&\textbf{6.580}&  \textbf{11.641}&\textbf{8.367}&\textbf{8.760} \\ 
			\cmidrule(lr){2-7} \cmidrule(lr){8-13} 
		&\multicolumn{13}{c} {Tropics}\\
				
				\cmidrule(lr){2-13} 
				&	\multicolumn{6}{c}{Absolute Percentage Error} &\multicolumn{6}{c}{Mean Scaled Interval Score}  \\
				&$h$=1 &$h$=2&$h$=4&$h$=6&$h$=1:3 &$h$=1:6&$h$=1&$h$=2&$h$=4 &$h$=6&$h$=1:3&$h$=1:6 \\
	\cmidrule(lr){2-7} \cmidrule(lr){8-13} 
        VARwT &44.422& 62.163&  69.043&  79.502&58.570&65.516&{6.140}& \textbf{9.145}&  14.144&  33.223&\textbf{8.718}&15.790  \\ 
        DeepAR & 52.113 &{48.542} & 46.185 & 60.936 & 46.192 & 50.176 & 40.178 & 34.033 & 31.429 & 39.313 & 31.952 & 33.657 \\ 
        DeepState &42.780 & 56.903 & 29.021 & 43.004 & 47.921 & 41.990 & 28.412 & 47.050 & 26.967 & 39.685 & 38.220 & 35.555 \\ 
 
        DeepVARwT &\textbf{29.205}&\textbf{41.079}& \textbf{28.770}&  \textbf{32.942}&\textbf{36.352}&\textbf{33.611}&\textbf{4.996}& 13.602&   \textbf{8.902}&  \textbf{10.032}&  9.865& \textbf{9.358}  \\ 
				\midrule
			\end{tabular}
		\end{adjustbox}
	\end{center}
	\end{table}
Overall, the DeepVARwT model gave better forecasts and prediction intervals than other models, especially for the Northern and Southern Hemisphere series. 

The number of weight parameters in the network for the 20 fitted DeepVARwT models varied between 444 and 508. The computation time for generating 20 sets of predictions using VARwT, DeepAR, DeepState and DeepVARwT was about 1, 45, 80, and 30 minutes, respectively.
\pagebreak
\subsection{US macroeconomics series 2}
\label{usmacro-bvarsv-sec}
\begin{figure}[H]
	\centering
	\includegraphics[width=0.7\linewidth]{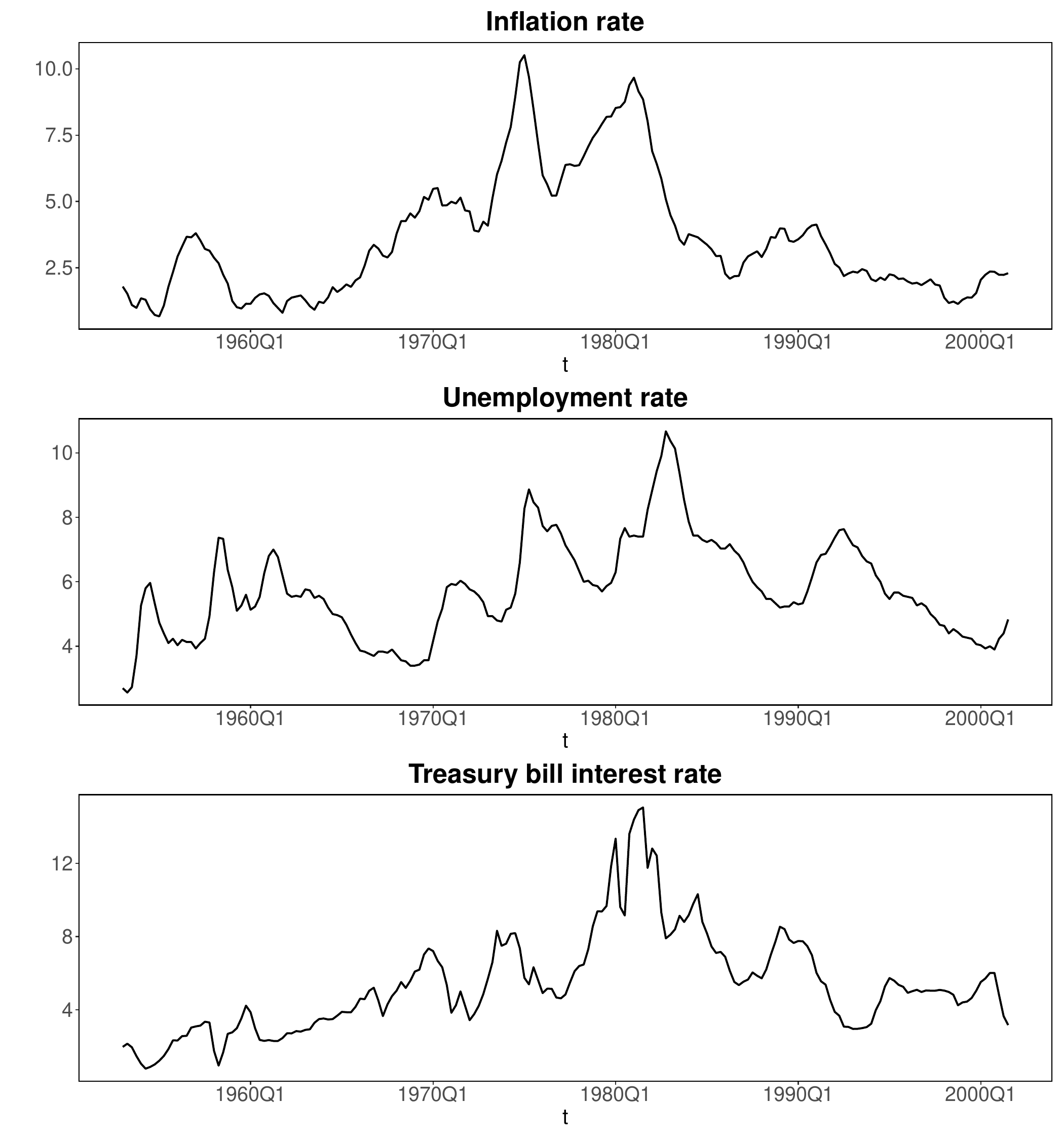}
	\caption{Inflation rate, unemployment rate and treasury bill interest rate for the US from 1953Q1 to 2001Q3} 
	\label{usmacro-bvarsv}
\end{figure}

We continue to apply our model to another set of US macroeconomic data (Fig.\ref{usmacro-bvarsv}) including inflation rate (year-over-year log growth rate of the GDP price index), unemployment rate and treasury interest rate from 1953Q1 to 2001Q3, as analysed by \cite{primiceri2005time}. {The inflation rate differs from the first real data example where it was defined as ``the percentage change in the GDP, chain-weighted price index at annual rate"  \cite{jorda2005estimation}}. 
Each series consists of 195 observations and exhibits a clear trend. 
We fitted a DeepVARwT(4) model to these series and  forecast $h=1,2,...,8$ steps ahead of time $T=168$. To be consistent with our previous real data applications, we repeated this 19 times, each time moving the training sample forward by one time point. 

The maximum power of $t$ was chosen among $2,3,4$ and the hidden state size out of $10,12,15$. We employed the learning rates  $\eta_{1}=0.0005$ and $\eta_{2}=0.01$, with $K=500$ iterations and precision $\mbox{\em prec}=10^{-7}$. The number of weight parameters in the network for the 20 fitted DeepVARwT models varied between 635 and 2,185.

The forecasts will be compared with those from a VARwT(4) model using $\mathbf{\color{black}z}_t=(t,t^2,\ldots,t^9)^{\prime}$ to account for the number of turning points in the series, DeepAR,
and DeepState 
models with default hyperparameters. The computation time to generate 20 sets of predictions using VARwT, DeepAR, DeepState and DeepVARwT was about 1, 45, 80, and 35 minutes, respectively.

\begin{figure}[h!]
	\centering
	\includegraphics[width=0.7\linewidth]{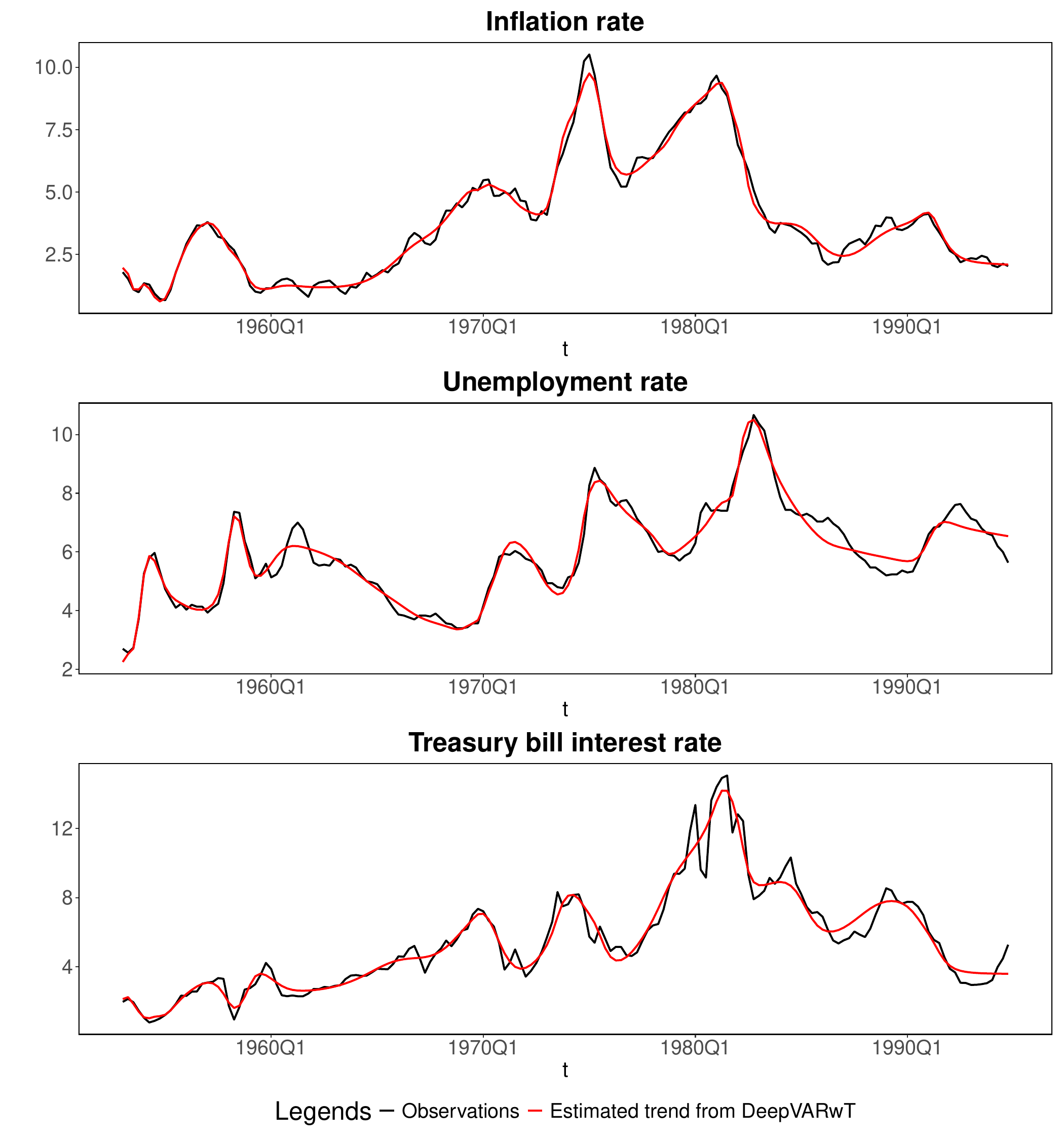}
			\vspace*{-5mm}
	\caption{The first training sample (black lines) from 1953Q1 to 1994Q4 and the corresponding estimated trends (red lines).} 
	\label{fig:estimated-trend-usmarco-bvarsv}
\end{figure}

\begin{figure}[h!]
	\centering
	\includegraphics[width=0.8\linewidth]{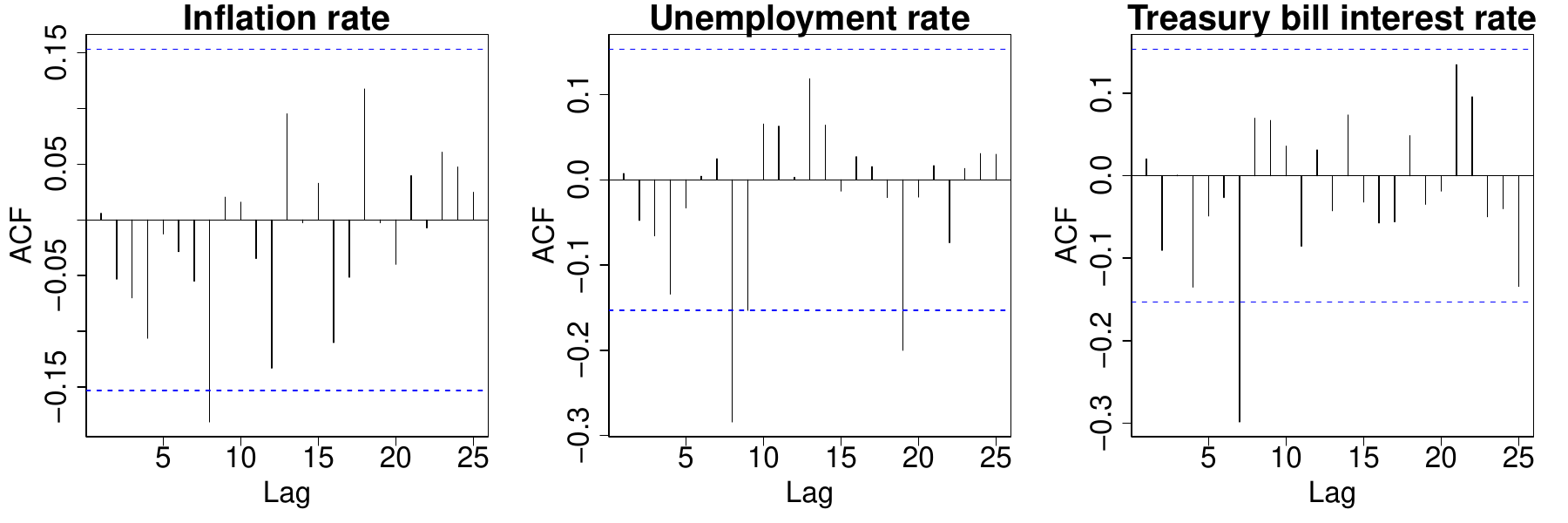}
			\vspace*{-5mm}
	\caption{Sample autocorrelations of residuals.} 
	\label{fig:acf-residuals-section0-usmacro-bvarsv}
\end{figure}

\FloatBarrier
From Fig.~\ref{fig:estimated-trend-usmarco-bvarsv}, we can see that the estimated trends (red lines) for the first training sample $(i=1)$ follow the observations (black lines) smoothly.
\FloatBarrier
The sample autocorrelations of residuals are shown in Fig.~\ref{fig:acf-residuals-section0-usmacro-bvarsv}. 
{The results are reasonably good for the inflation rate series and the treasury bill interest rate series with one value outside the boundaries, while the unemployment rate series had two.}

\FloatBarrier

Fig.~\ref{fig:qq-plots-climate} contains normal QQ plots of the residuals. There is some deviation from normality for
all the series at both ends.

\begin{figure}[h!]
	\centering
	\includegraphics[width=0.7\linewidth]{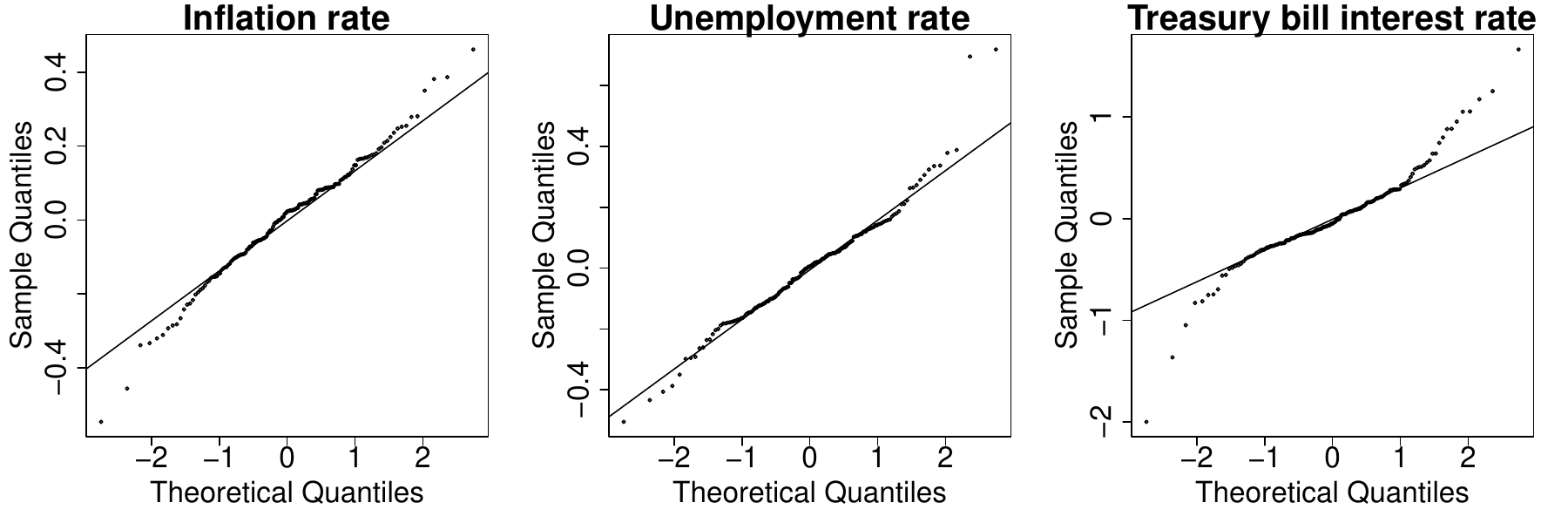}
			\vspace*{-5mm}
 \caption{Normal QQ plots of residuals.}
  \label{fig:qq-plots-usmacro-bvarsv}
\end{figure}
\FloatBarrier
Table \ref{mean-accuracy-climate} shows the APE and SIS values of different models at several horizons $h=1,2,4,8$ and averaged over $h=1:4$ and $h=1:8$. 

\begin{itemize}
	\tightlist

		\item \textbf{DeepVARwT vs VARwT.} Compared with the time-invariant VAR with trend, DeepTVARwT produced better point forecasts at almost all forecasting horizons for all the series (except $h=1$ for inflation rate and $h=1, 2$ for treasury bill interest rate).
		It also gave more accurate prediction intervals at almost all forecasting horizons for all the series (except $h=1, 2$ for inflation rate). 
	\item \textbf{DeepTVARwT vs other deep learning based models.} Compared with DeepAR and DeepState, our model produced more accurate point forecasts at almost all forecasting horizons for unemployment rate (except $h=1$) and treasury bill interest rate (except $h=1, 2$). Our model resulted in better prediction intervals at all forecasting horizons for unemployment rate and treasury bill interest rate.

\end{itemize}

	\begin{table}[h]
	\begin{center}
		\normalsize
		\centering
		\caption{Performance of DeepVARwT against other models according to APE and SIS.}
		\label{mean-accuracy-uc-macro}
		\begin{adjustbox}{width=1\textwidth}
			\begin{tabular}{rrrrrrrrrrrrrrrrr}
				\toprule
				&\multicolumn{13}{c} {Inflation rate}\\
				
				\cmidrule(lr){2-13} 
				&	\multicolumn{6}{c}{Absolute Percentage Error} &\multicolumn{6}{c}{Scaled Interval Score}  \\
			&$h$=1 &$h$=2&$h$=4&$h$=8&$h$=1:4 &$h$=1:8&$h$=1&$h$=2&$h$=4 &$h$=8&$h$=1:4&$h$=1:8 \\
	\cmidrule(lr){2-7} \cmidrule(lr){8-13} 
        VARwT & 9.961  &24.991  & 76.114  &259.015&39.129 &111.141& \textbf{1.281}& \textbf{2.388}&   14.168&  139.702& 5.851& 42.729\\ 
        DeepAR &\textbf{7.763}& \textbf{15.280}& \textbf{28.017}& 52.398& \textbf{18.497}& 31.416& 1.503&  3.558&{9.151}&  19.145&\textbf{5.164}&10.601 \\ 
        DeepState & 18.381& 22.345&  29.072&{40.687}& 24.188& {31.295}&5.416&  7.930&  10.453& 14.496& 8.176& 11.146\\ 
        DeepVARwT &13.607& 23.723& 34.295&  \textbf{33.111}& 25.267& \textbf{29.095}& 2.001&  4.480&    \textbf{8.940}&   \textbf{11.295}&{5.417}& \textbf{7.926}\\ 
        	\cmidrule(lr){2-7} \cmidrule(lr){8-13} 
       	&\multicolumn{13}{c} {Unemployment rate}\\
				
				\cmidrule(lr){2-13} 
				&	\multicolumn{6}{c}{Absolute Percentage Error} &\multicolumn{6}{c}{Scaled Interval Score}  \\
				&$h$=1 &$h$=2&$h$=4&$h$=8&$h$=1:4 &$h$=1:8&$h$=1&$h$=2&$h$=4 &$h$=8&$h$=1:4&$h$=1:8 \\
	\cmidrule(lr){2-7} \cmidrule(lr){8-13} 
        VARwT &3.514&  8.960&  25.229&  93.339& 13.425& 38.390&{1.263} & 2.738&   15.992&   139.895& 6.323& 45.909 \\ 
        DeepAR &\textbf{3.250}&  6.524& {13.014}& 31.404&8.099&16.310&1.273&  2.965&    8.728& 21.958&4.659&10.335 \\ 
        DeepState &15.111& 15.968&  19.593&  25.698& 17.509& 20.591&6.296&  9.376&  11.440& {15.244}& 9.717& 12.364\\ 
        DeepVARwT &3.269&  \textbf{5.578}&    \textbf{9.888}& \textbf{18.805}& \textbf{6.570}& \textbf{10.977}& \textbf{1.243}&   \textbf{2.413}&   \textbf{5.460}&  \textbf{15.137}& \textbf{3.119}& \textbf{6.877}\\ 
	        	\cmidrule(lr){2-7} \cmidrule(lr){8-13} 
	            	&\multicolumn{13}{c} {Treasury bill interest rate}\\
				
				\cmidrule(lr){2-13} 
				&	\multicolumn{6}{c}{Absolute Percentage Error} &\multicolumn{6}{c}{Scaled Interval Score}  \\
				&$h$=1 &$h$=2&$h$=4&$h$=8&$h$=1:4 &$h$=1:8&$h$=1&$h$=2&$h$=4 &$h$=8&$h$=1:4&$h$=1:8 \\
	\cmidrule(lr){2-7} \cmidrule(lr){8-13} 
        VARwT &  5.390& 10.537&  20.041&  75.861& 12.581& 30.385& 2.110&  {3.211}&   4.875&   44.500& 3.452& 13.042 \\ 
        DeepAR &\textbf{4.868}& \textbf{10.073}& 18.693& 24.993&12.340&18.016&{1.583} & 3.904&   13.264&  18.203& 6.848& 12.436\\ 
        DeepState &13.111& 13.448&  14.801& 22.705& 13.860& 16.391&3.135&  3.869&    5.363&  10.175& 4.191& 6.191\\ 
        DeepVARwT &7.169& 11.176&  \textbf{12.521}& \textbf{18.190}& \textbf{10.498}& \textbf{13.326}& \textbf{1.409}&  \textbf{2.673}&    \textbf{3.076}&  \textbf{5.054}& \textbf{2.440}& \textbf{3.437}\\
				\midrule
			\end{tabular}
		\end{adjustbox}
	\end{center}
	\end{table}
\FloatBarrier
\vspace*{-6pt}
Over $h=1:4$ and $h=1:8$, the DeepVARwT model outperformed other models in providing better forecasts and prediction intervals, except for the inflation rate over h=1:4 where DeepAR did better. 

The number of weight parameters in the network for the 20 fitted DeepVARwT models varied between 849 and 1350. The computation time for generating 20 sets of predictions using VARwT, DeepAR, DeepState and DeepVARwT was about 1, 45, 80, and 59 minutes, respectively.

\section{Summary and further discussion}
\label{conclusion-varwt}
In this work, we proposed a new approach to VAR modelling and forecasting by generating trends as well as model parameters using an LSTM network and the associated deep learning methodology for exact maximum likelihood estimation. A simulation study demonstrated the effectiveness of the proposed approach and showed uncertainty in the estimates in terms of bias, standard deviation and mean square error. 
Three examples with real data are provided to show that it competes well with existing models in terms of prediction performance.  

The order of autoregression was set equal among the different models fitted to real data. A reasonably high order of polynomial trend was used in VARwT. Default values of the hyper-parameters (2  hidden layers each of 40 units etc) for the DeepAR and DeepState models were used, which worked well for the model fitting and thus not tweaked for better performance. The future values were only used for comparisons with predictions from each fitted model. The \code{python} code is available at \url{https://github.com/lixixibj/DeepVARwT-data-code} for reproducing the forecasting results.

The computation becomes more challenging as the number/length of the component series increases. With high dimensional time series, a potential avenue for future research involves incorporating regularisation and low-rank structure into the model fitting. One approach is to impose a low-rank assumption on $A_1,...,A_p$ combined into a single matrix. This enables a reduction along one specific direction \citep{wang2022high}.
Building on this concept, \cite{wang2022high} further rearranged $A_i, i=1,...,p$ into a tensor to reduce the dimension along three directions, allowing each direction to have a different low-rank structure.  Incorporating a tensor structure into our DeepVARwT model could be a potential future direction.

{
The model that has been explored so far rely on the assumption of Gaussianity. However, in practical applications such as demand forecasting, series may exhibit sporadic occurrences with periods of no activity at all. This intermittent behaviour of demand calls for the relaxation of the Gaussian assumption to accommodate discrete data. It is possible to generalise the digitised Gaussian ARMA model of \cite{lennon2019estimation} to the multivariate case.}

\vspace*{-2pt}
\section*{Acknowledgments}

The first author's work was supported by The University of Manchester under a Dean's Doctoral Scholarship Award. {\color{black}We would like to thank all three referees and the Associate Editor for their questions and comments that helped improve the presentation of this paper.}
\section*{Disclosure statement}
No potential conflict of interest was reported by the authors. 

\vspace*{-4pt}
\bibliographystyle{tfs}
\bibliography{DeepVARwT}

\appendix
\section{Prediction error variances and covariances}
\label{appendix}

First consider the model for $\{\textbf{y}_{t}\}$ to be VAR(1) with trend:
\begin{equation}
\label{var1}
\textbf{y}_{t}-\bm{\mu}_{t}=A(\textbf{y}_{t-1}-\bm{\mu}_{t-1})+\bm{\varepsilon}_{t},
\end{equation}
where $\{\bm{\varepsilon}_{t}\}$ is white noise, $\bm{\varepsilon}_{t} \sim \mathcal{N}(\bm{0},\,\Sigma)$ and $\bm{\varepsilon}_{t}$ is uncorrelated with $\textbf{y}_{t-1}, \textbf{y}_{t-2},....$.

Then, we can decompose $\textbf{y}_{T+\ell}$ starting with $\textbf{y}_{T}$  for $\ell=1,...,h$:
\begin{equation}
\label{decomp}
\begin{aligned}
\textbf{y}_{T+1}-\bm{\mu}_{T+1}&=A(\textbf{y}_{T}-\bm{\mu}_{T})+\bm{\varepsilon}_{T+1}, \\
\textbf{y}_{T+2}-\bm{\mu}_{T+2}&=A(\textbf{y}_{T+1}-\bm{\mu}_{T+1})+\bm{\varepsilon}_{T+2}=A^{2}(\textbf{y}_{T}-\bm{\mu}_{T})+A\bm{\varepsilon}_{T+1}+\bm{\varepsilon}_{T+2}, \\
\textbf{y}_{T+3}-\bm{\mu}_{T+3}&=A^{3}(\textbf{y}_{T}-\bm{\mu}_{T})+A^{2}\bm{\varepsilon}_{T+1}+A\bm{\varepsilon}_{T+2}+\bm{\varepsilon}_{T+3}, \\
&\vdots \\
\textbf{y}_{T+h}-\bm{\mu}_{T+h}&=A^{h}(\textbf{y}_{T}-\bm{\mu}_{T})+\sum_{i=0}^{h-1}A^{i}\bm{\varepsilon}_{T+h-i},
\end{aligned}
\end{equation}
where $A^{i}$ is understood to be the identity matrix when $i=0$.

From~(\ref{decomp}), the best linear predictor for $\textbf{y}_{T+\ell}$ given $\mathbf{y}_{T}, \mathbf{y}_{T-1}, \ldots$ is
\begin{equation}
\label{b-l-p}
\hat{\mathbf{y}}_{T+\ell}=\mbox{E}\left[\mathbf{y}_{T+\ell} \mid \mathbf{y}_{T}, \mathbf{y}_{T-1}, \ldots\right]=A^{\ell}(\textbf{y}_{T}-\bm{\mu}_{T})+\bm{\mu}_{T+\ell},
\end{equation}
and the associated prediction error variance-covariance matrix is
\begin{equation}
\label{preidtion-error}
\mbox{Var}\{\mathbf{y}_{T+\ell}-\hat{\mathbf{y}}_{T+\ell}\}=\sum_{i=0}^{\ell-1}A^{i} \Sigma(A^{i})^{\prime}.
\end{equation}

When $\{\textbf{y}_{t}\}$ follows the VAR($p$) model~(\ref{varwt}) with trend, we use its VAR(1) form
\begin{equation}
\textbf{y}_{t}^{\ast}-\bm{\mu}_{t}^{\ast}= A^{\ast}(\textbf{y}_{t-1}^{\ast}-\bm{\mu}_{t-1}^{\ast})+\bm{\varepsilon}_{t}^{\ast},
\end{equation}
where $\textbf{y}_{t}^{\ast}=(\textbf{y}_{t}^{\prime},\textbf{y}_{t-1}^{\prime},...,\textbf{y}_{t-p+1}^{\prime})^{\prime}$, $\bm{\mu}_{t}^{\ast}=(\bm{\mu}_{t}^{\prime},\bm{\mu}_{t-1}^{\prime},...,\bm{\mu}_{t-p+1}^{\prime})^{\prime}$,
 \begin{equation}
  \begin{aligned}
A^{\ast}=\left[\begin{array}{ccccc}
A_{1} & A_{2} & \cdots & \cdots &A_{p} \\
I & \text{\Large o} & \cdots & \cdots & \text{\Large o} \\
\text{\Large o} & I & \ddots & & \vdots \\
\vdots & \ddots & \ddots & \ddots & \vdots \\
\text{\Large o} & \cdots & \text{\Large o} & I & \text{\Large o}
\end{array}\right],
  \end{aligned}
\end{equation}
and 
 \begin{equation}
  \begin{aligned}
\mathbf{y}_{t}=[I,\text{\Large o}, \cdots, \text{\Large o}]\textbf{y}_{t}^{\ast}
  \end{aligned}.
\end{equation}
The variance-covariance matrix of ${\bm{\varepsilon}_{t}^{\ast}}=({\bm{\varepsilon}}_{t}^{\prime}, \textbf{0}^{\prime},...,\textbf{0}^{\prime})^{\prime}$ is
 \begin{equation}
  \begin{aligned}
\Sigma^{\ast}=\left[\begin{array}{cc}\Sigma& \text{\Large o}\\ \text{\Large o}& \text{\Large o}\end{array}\right].
  \end{aligned}
\end{equation}
Using~(\ref{b-l-p}), the best linear predictor for $\mathbf{y}_{t+h}^{\ast}$  given $\textbf{y}_{t}^{\ast},\textbf{y}_{t-1}^{\ast},... $ is
\begin{equation}
\label{best-linear-predictor}
\hat{\mathbf{y}}_{t+h}^{\ast}=({A}^{\ast})^{h}(\mathbf{y}_{t}^{\ast}-\bm{\mu}_{t}^{\ast})+\bm{\mu}_{t+h}^{\ast}.
\end{equation}
Using~(\ref{preidtion-error}), the variance-covariance matrix of the prediction error for  $\hat{\mathbf{y}}_{t+h}^{\ast}$ is
\begin{equation}
\label{var-cov}
\sum_{i=0}^{h-1} ({A}^{\ast})^{i} \Sigma^{\ast}(({A}^{\ast})^{i})^{\prime}.
\end{equation}
The prediction for $\mathbf{y}_{t+h}$ can be extracted from that for  $\mathbf{y}_{t+h}^{\ast}$. 
The prediction error variance-covariance matrix for $\hat{\textbf{y}}_{t+h}$ is in the top-left corner of the above.

\end{document}